\documentclass[12pt]{article}
\usepackage[square,numbers]{natbib}
\usepackage[utf8]{inputenc}
\usepackage[T1]{fontenc}
\usepackage{graphicx}
\usepackage{lscape}

\title{Modulation of oculomotor control during reading of mirrored and
inverted texts}

\author{Johan Chandra, Andr\'{e} Kr\"{u}gel, Ralf Engbert \\
	Deparment of University of Potsdam, Potsdam, Germany }
	
\date{\today}
\begin{document}
\maketitle

\begin{abstract}
The interplay between cognitive and oculomotor processes during reading can be explored when the spatial layout of text deviates from the typical display. In this study, we investigate various eye-movement measures during reading of text with experimentally manipulated layout (word-wise and letter-wise mirrored-reversed text as well as inverted and scrambled text). While typical findings (e.g., longer mean fixation times, shorter mean saccades lengths) in reading manipulated texts compared to normal texts were reported in earlier work, little is known about changes of oculomotor targeting observed in within-word landing positions under the above text layouts. Here we carry out precise analyses of landing positions and find substantial changes in the so-called launch-site effect in addition to the expected overall slow-down of reading performance. Specifically, during reading of our manipulated text conditions with reversed letter order (against overall reading direction), we find a reduced launch-site effect, while in all other manipulated text conditions, we observe an increased launch-site effect. Our results clearly indicate that the oculomotor system is highly adaptive when confronted with unusual reading conditions.
\end{abstract}

\thispagestyle{empty}

\section{Introduction}

Visual acuity is greatest at the center of visual field (the {\sl fovea}) and declines sharply on the periphery, which limits the information extraction process of visual input from the environment. To compensate the limitation, the eyes generate short and rapid movements, {\sl saccades}, to shift the fovea to the regions of interests for high-acuity information processing \citep{FindlayGilchrist2003}. Similar principle is observed during processing of writtten text or reading. During reading, the eyes typically move forward about 6-7 character spaces during saccades and fixate on a word for about 200-250~ms to support word processing. The control of (saccadic) eye movements requires the coordination of several fundamental cognitive subsystems such as word recognition, attention \cite{SWIFT2005}, and oculomotor control \citep{Reicheleetal1999}. While the cognitive system is responsible for selecting which word to be fixated next, it is the oculomotor system that is responsible for shifting the fovea to the regions of interests for high-acuity information processing \citep{FindlayGilchrist2003}. Thus, our reading ability depends on the oculomotor performance, whose properties are reflected most clearly in the statistics of within-word fixations, typically the eyes' landing position on words after saccades. 

Unlike the well-documented effects of cognitive modulation on temporal aspects (e.g. fixation duration) of eye movement measures \citep[e.g.][]{Kliegletal2004,Rayner2009}, small effects of cognitive modulation on spatial aspects (e.g. within-word landing position) of eye movement measures were reported. For example, orthographic familiarity and regularity influence landing positions \citep{Hyona1995,WhiteLiv2004, WhiteLiv2006b, WhiteLiv2006a}. Furthermore, corpus analyses showed a significant effect of word frequency on mean fixation position: Saccades landed further into the (3- to 6-letter) target word, when it was a high-frequency word as compared to a low-frequency word \citep{Nuthmann:2006dis}. Lavigne \cite{Lavigne2000} reported a shift of initial fixation location toward the end of high predictable words, but only when the words were seen more frequently and for saccades lauched near the word beginning. On the other hand, Rayner \cite{Rayner2001} reported that word predictability had little influence on initial landing position on word and suggested that landing position effects in reading were primarily modulated by low-level processing. Finally, data from z-string scanning (where all letters were replaced by the letter ``z'' or ``Z'') indicated that within-word landing position distributions are very stable and do not critically depend on meaningful content (\cite{Nuthmann2007}; see also \cite{LukeHenderson2013}). Moreover, most effects of higher-level processing on mean fixation position are small ($<0.5$ character spaces).

Interestingly, the within-word landing position were reported to influence the "higher-level" processes. In several studies, it was reported that (isolated) word recognition time was at the minimum when the eyes land at the center of the word compared to when the eyes land on the word's periphery, termed the \textit{optimal viewing position} \citep{OReganLevySchoen1987,OReganOVP1984}. Similar but weaker effect was observed on refixation probability in reading: readers were less likely to refixate the words if fixation land near the word center \citep{VituOVP}. Oppositely, mean single fixation duration is the greatest when the eyes land at word center, termed the \textit{inverted optimal viewing position} in Vitu et al. \cite{VituIOVP}. To explain the effect, Nuthmann and colleagues \cite{Nuthmannetal2005} argued that fixations landed on the word egdes were typically not intended to land on the target (mislocated fixations), reflected on the short durations.

The observed evidences in reading researches are the result of the integration of cognitive processes in word selection and oculomotor processes in shifting the eyes to the area of interest. To isolate the underlying process affected the observation, one should manipulate factors associated to one process while controling the remaining factors.  In fact, Kolers and Perkins \cite{KolersPerkins1975} used geometric rotations, reflections and other transformations of text as the physical variation to study the recognizability of the texts and the influence that practice in reading one type of transformation applied on the recognition of others. They found that the transformations being tested varied in difficulty and transferability. Likewise, Kowler and Anton \cite{KowlerAnton1987} applied similar types of transformation to test the effects on global saccade lengths and fixation durations. By observing eye movement patterns of two participants, they reported that the directional pattern of saccades had relatively modest effects on reading speed under the instruction to read accurately. They argued that the reading time was affected by longer time needed to generate short saccades observed in reading difficult texts. In a separate test, they found negative relationship between saccade length and saccade latency: short saccades (less than 30') have longer latency than long saccades. Additionally, as a response to the Internet myth, Rayner et al. \cite{Rayner2006} tested different types of word transposition (internal, beginning, and end of word) on reading time and reported that although participants were able to read the text, reading time was slower for some transposition types, especially when the word beginning was transposed. Hence they concluded the importance of word beginning (see also \cite{Whiteetal2008} for similar conclusion). Following the above approach of presenting texts in unfamiliar representation, we designed a study with four different experimental conditions and a control condition to systematically investigate the possible modulations of oculomotor processes in response to variations of the spatial layout of texts. Furthermore, the current study employs various eye movement measures to describe the oculomotor performances during reading.

\begin{figure}[p]
\unitlength1mm
\begin{picture}(150,145)
\put(10,75){\includegraphics[width=120mm]{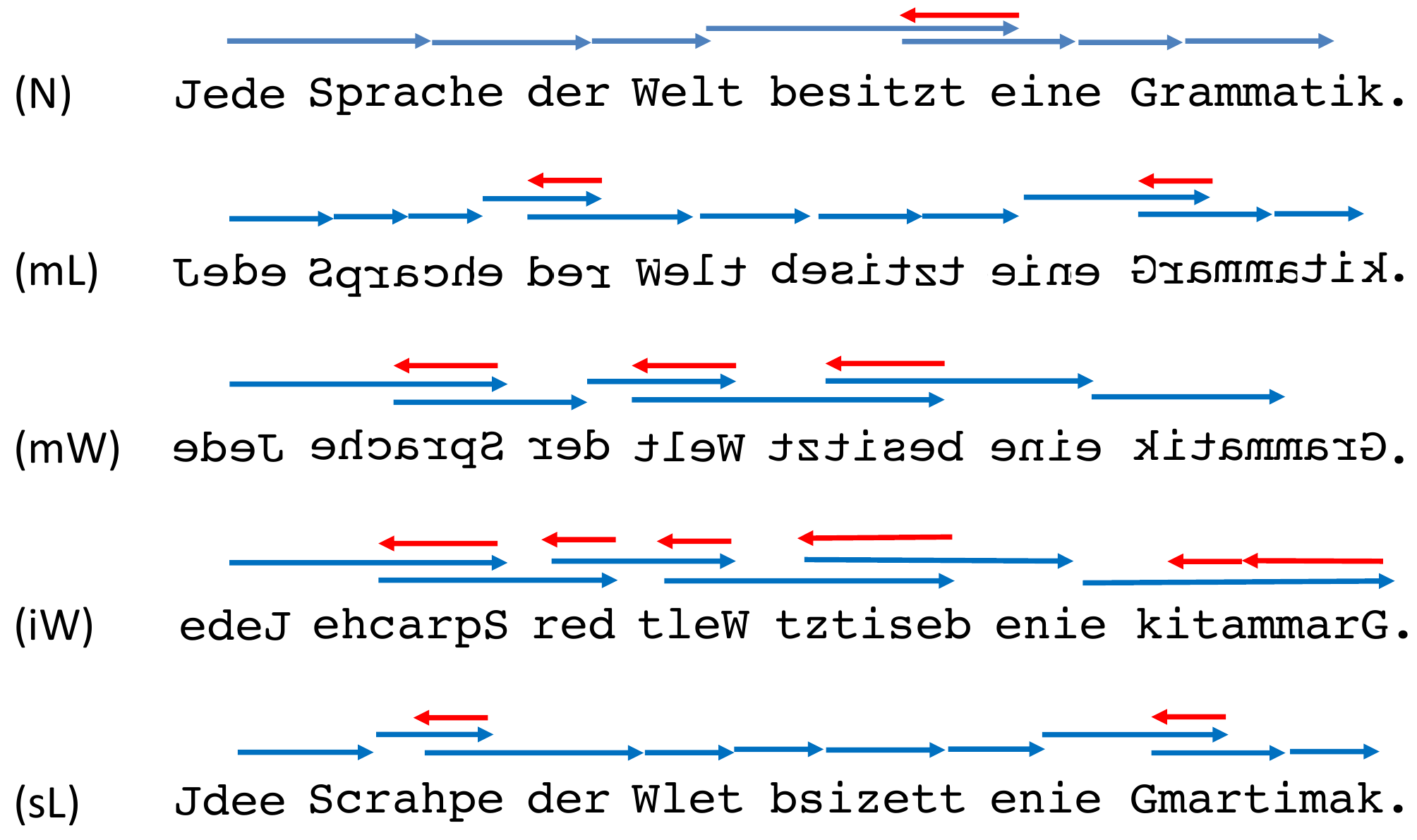}}
\put(-8,0){\includegraphics[width=152mm]{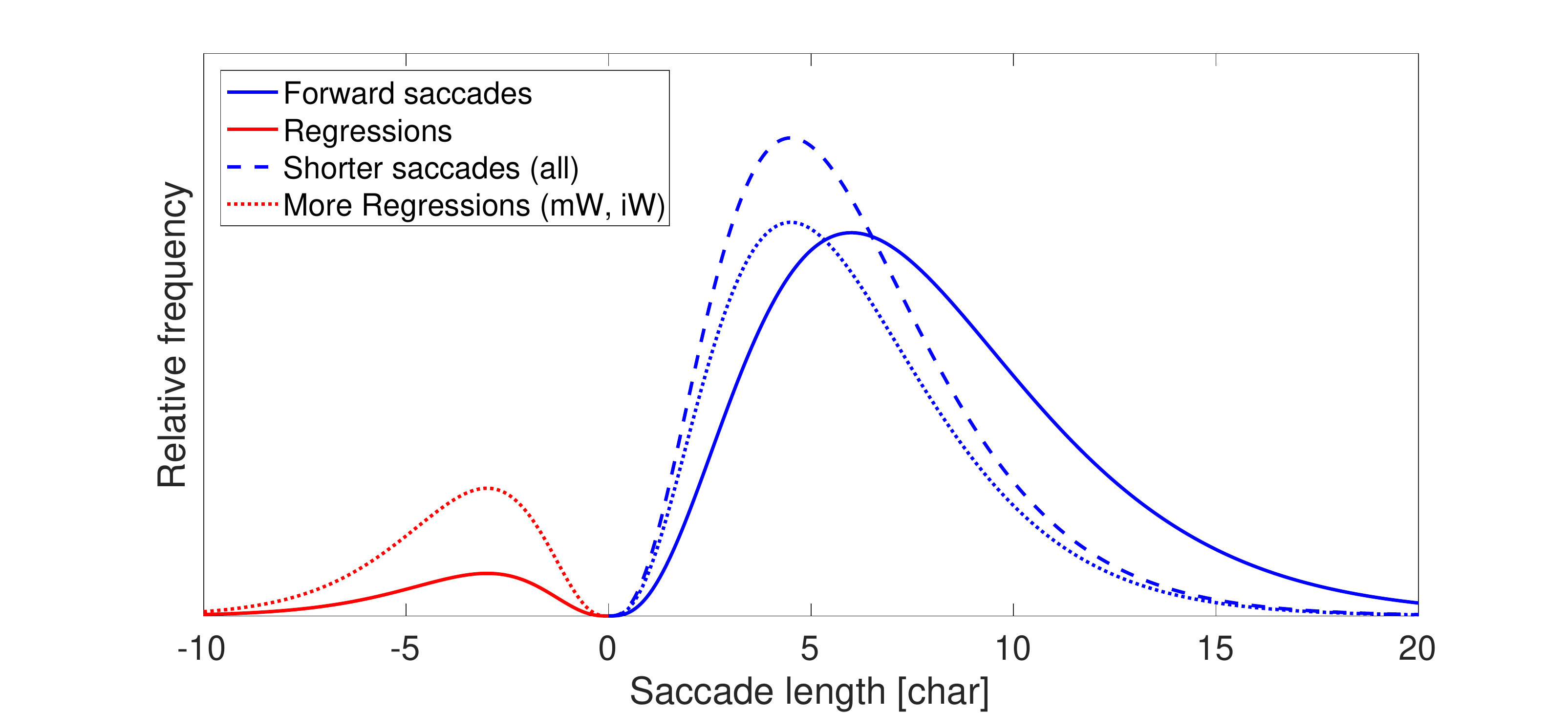}}
\put(0,0){(b)}
\put(0,76){(a)}
\end{picture}
\vspace{0mm}
\caption{\label{fig_hypos}
Experimental sentence stimuli and hypotheses on saccade lengths. (a)  In the control condition, normal German text (N) is presented; for mirrored letters (mL) and scrambled words (sL), we expect shorter saccade lengths, on average. For mirrored words (mW) and inverted words (iW), word beginnings and word ends are exchanged, so that we expect longer mean saccade lengths and more regressions due to more frequent re-readings of the same string. (b) The distribution of saccade lengths (solid lines) and the expected changes due to the experimental manipulations.
}
\end{figure}

In current study, letter positions and word representations were experimentally manipulated in the following ways: We used texts composed of mirror-inverted letters (mL), mirror-inverted words (mW), inverted words (iW), where regular letters are printed in reverse order, and scrambled letters (sL)  (see Figure~\ref{fig_hypos}a for an example). In mirrored-words (mW) and mirrored-letter (mL) conditions, either the complete word or the constituting letters were mirror-inverted with respect to the vertical axis. In contrast, no mirroring was involved in the construction of inverted-word (iW) and scrambled-word (sL) text conditions. In inverted words (iW) condition, letter representation was normal, but the position was inverted in the iW condition to mimic the letter position in the mW condition. In scrambled letters (sL) condition, the positions of the first and the last letter of a word were maintained, while the letter positions in between were randomized (i.e., there was no change in words with length less than 4 letters). The condition mL, mW and iW are equivalent to the condition NNV, NRV and NRN conditions in \cite{KowlerAnton1987} while the condition sL is equivalent to the internal transposition manipulation in \cite{Rayner2006}.

The following sections describe the robust finding on within-word landing position distribution during reading and the proposed models to explain the observed phenomena. Furthermore, we will discuss about current reading models and their predictions, particularly on saccade generation and "where" the eyes land, in relation to the manipulations in current study.

\subsection*{Within-word landing positions}
Regarding where the eyes land during reading, a robust finding is that within-word landing position approximately follows a Gaussian density function, with a pronounced peak, typically located halfway between word beginning and word center, but with a surprisingly large variance \citep{Rayner1979,McConkieetal1988,Oregan1990}. The landmark study by McConkie et al. \cite{McConkieetal1988} identified two independent oculomotor error components in reading, which we will denote as the {\sl range-error model} throughout this article: (i) The {\sl random placement error} is assumed to reflect perceptuo-oculomotor inaccuracy in the execution of saccades, which can be approximated by a Gaussian distribution. (ii) The {\sl saccadic range error} represents a systematic, launch-site contingent shift of mean landing positions and is typically explained as a general response bias of the human motor system \citep{poulton1981}. 

Specifically, McConkie and colleagues \cite{McConkieetal1988} found that during reading, the within-word landing positions varied systematically with the launch-site distance, i.e., the distance between fixation location (before the saccade) and the beginning of the target word. The stable observation in reading is that each letter increment of the saccade launch-site generates a shift of mean landing-site with a magnitude of about  half a character space to the left, the {\sl launch-site effect}, which is independent of the target word length. If the distances between launch-sites and landing sites are measured relative to word centers, then the within-word mean landing position can be described by a linear landing-position function \citep[see also][]{RadachMcConkie1998} of the form
\begin{equation}
\label{eq_launch_site_effect}
\Delta_{PVP} = \lambda\cdot (L_0-L) \;,
\end{equation}
where $\Delta_{PVP}$ denotes the average shift of the within-word mean landing position from the word center and $L$ is the distance between launch site and the center of the target word. While a negative value of $\Delta_{PVP}$ indicates that the within-word mean landing position shifts to the left of the word center, a positive value indicates a rightward shift. The parameter $L_0$ represents the center-based launch-site distance, where saccades land precisely on the word center, on average. The strength of the launch-site effect is represented by the slope parameter $\lambda$. An estimated slope of $\lambda\approx 0.5$ was observed in readers of English \citep{McConkieetal1988} and German \citep{Nuthmannetal2005} texts. 

While the range-error model was generally a successful first description of the eye-movement data, experimental studies demonstrated effects that could not be explained within this model. First, the presence of additional stimuli could influence the saccadic landing positions \citep{Coeffe1987,Deuble1984,Findlay1982,Vitu1991,Vitu2008,Vituetal2006}. Second, Krügel and Engbert \cite{KruegelEngbert2010} demonstrated that saccade type (i.e.~word skipping) could influence the saccade landing positions during reading \citep[see also][]{Kruegeletal2012,Radach1996}. These findings challenged the range-error model in explaining observed saccadic landing positions during reading. 

It is important to note, however, that the range-error model is purely descriptive, since it does not include more fundamental computational principles for oculomotor control. Furthermore, the slope parameter of $\lambda$ represents quantification of the strength of the launch-site effect, without direct inferences on what processes underlying the observed phenomenon. As a consequence, it is not surprising that integrating new experimental evidence in the range-error model is difficult. In the next section, we discuss a process-oriented Bayesian model of within-word fixation position that was developed over the last 10 years.

\subsection*{A Bayesian model of oculomotor control in reading}
The framework of Bayesian decision theory has been proposed as a principled approach to the optimal control of human behavior in the context of integration of sensorimotor and cognitive processes \citep{Koerding2007,KoerdingWolpert2004,KoerdingWolpert2006}. Since saccadic eye movements require both sensorimotor (i.e., moving the eyes to foveate words) and cognitive processes, Engbert and Krügel \cite{EngbertKruegel2010} proposed that eye movement control during reading could be explained using Bayesian estimation. According to Bayes rule \citep[e.g.,][]{Wolpert2005}, the optimal estimate of a target position $x$, given a sensory observation at $x_0$, can be calculated as the conditional probability (posterior)
\begin{equation}
\label{eq_bayes}
\pi(x|x_0) \sim q(x_0|x)p(x) \;,
\end{equation}
where $p(x)$ is the previously learned prior distribution of the target, independent of current sensory input, and the conditional probability $q(x_0|x)$ is the sensory likelihood of the observation at position $x_0$ given a target at $x$. The relation in Eq.~(\ref{eq_bayes}) determines the dependence of the posterior probability from $x$. The missing constant of proportionality can be obtained by normalization.

Engbert and Kr\"ugel \cite{KruegelEngbert2010} proposed that the dependence of landing positions within a word on the saccadic launch site is a special case of the Bayesian principles, Eq.~(\ref{eq_bayes}), for saccade planning in reading. In a mathematical model, the likelihood $q(x_0|x)$ was modeled as an unbiased, normally distributed probability density centered at the intended target word and with variance $\sigma_0^2$, which represents the degree of sensory uncertainty. Assuming that the prior distribution is also normally distributed, the product of the two Gaussian densitiy functions results in another normal density function, the posterior probability density $\pi(x|x_0)$. The posterior probability provides a natural explanation of the launch-site effect, since the position of its maximum falls between the maximum of the prior distribution and the maximum of the sensory likelihood. As a result, the posterior reproduces the systematic tendencies of saccades (i) to overshoot the center of close target words and (ii) to undershoot the center of distant target words \citep{EngbertKruegel2010}. 

The shift of the mean of the posterior $\mu_P$ from the observation $x_0$ can be calculated as 
\begin{equation}
\label{eq_bayes_shift}
\Delta_{Bayes}=\mu_P-x_0 = \frac{\sigma_0^2}{\sigma_0^2+\sigma_T^2}(\mu_T-x_0) \;,
\end{equation}
where $\sigma^2_T$ is the variance of the prior probability of center-based launch-site distances. Comparing the equation for the launch-site effect, Eq.~(\ref{eq_launch_site_effect}), with the predicted effect in the Bayesian theory, Eq.~(\ref{eq_bayes_shift}), thus assuming $\Delta_{PVP}\equiv\Delta_{Bayes}$,we obtain
\begin{equation}
\lambda = \frac{\sigma_0^2}{\sigma_0^2+\sigma_T^2} \;.
\end{equation}

Recently, Krügel and Engbert \cite{KruegelEngbert2014} introduced an advanced model, which includes an explicit model for the computation of the word center from sensory estimates of word boundaries. Therefore, Bayesian models provide a robust theoretical framework to explain where the eyes move during reading. One advantage to model within-word landing position distribution using an explicit Bayesian model is that the interpretability of the results. The slope parameter $\lambda$ estimated from Bayesian model represents weighting of optimal behavior during reading: maintaining constant saccade length while targeting word center. For extreme case where $\lambda\to 0$, it can be interpreted that the optimal oculomotor control in reading puts more weight on the precision in landing on target precision, hence minimizing range error. On the other hand, the value of $\lambda\to 1$ means that optimal reading behavior rely on maintaining constant saccade length, reducing the importance of target location. With the typical empirical value of $\lambda\approx 0.5$, we obtain the important result that the sensory variance of the target location is approximately the same as the variance of the prior distribution, i.e., $\sigma_0^2\approx\sigma_T^2$, in optimal reading behavior.

\subsection*{Hypothesis and predictions from various reading models}
Reading models were typically developed to help understanding the complex processes underlying reading processes. Despite the fact that eye movement control during reading requires both cognition and oculomotor systems, as mentioned above, empirical studies found that cognition had small effects on within-word landing positions. Interestingly, adding additional visual cue \citep{Nuthmann2006} or changing sentence presentation, i.e. texts are read from top to bottom \citep{JohnsonStarr2017} did not notably change the landing position distributions on words. Consequently, most mathematical models of saccade generation during reading assume that oculomotor control is dissociated from cognitive processes. Cognitive-based reading models (e.g., E-Z Reader: \cite{Reichleetal1998,Reichleetal2006,Reichleetal2009};  SWIFT: \cite{Engbert2002,SWIFT2005,SchadEngbert2012}) assume that cognitive processes related to language processing are responsible for eye movements without distinguished effects on oculomotor within-word targeting process. Moreover, most cognitive models based their implementation of saccadic errors on the range-error model (\citep{McConkieetal1988}) with relatively fixed values. Since the manipulation types tested in current study maintain the spatial information such as word length, cognitive-based models will not predict substantial changes in within-word landing positions since they asumme that oculomotor process is mainly affected by ``low level'' information. However, since text manipulations will increase processing loads, hence affecting saccade generating time, these models will generate more refixation saccades but less skipping saccades. 

Most relevant to current study is Mr.Chips \citep{MrChips1997,MrChips2002}, an ideal-observer model that combine visual, lexical, and oculomotor information optimally to read simple texts in the minimum number of saccades. The model operates according to an entropy-minimization principle, generating saccades that minimize uncertainty about the current word or saccades that move the visual span furtherst to the right. Note that for Mr.Chips, a word is said to be fixated if the central slot of the visual span (a linear array of character slots with each slot can be either high or low resolution) falls on one of the letters of the word and this central slot does not have preferred status. Mr. Chips' skipping rate and global landing position distribution were similar to human data. Mean saccade length decreases as the lexion size increases. Furthermore, it generated less refixations and reduced launch-site effect ($\lambda = 0.21$) in reading normal text. Given that word identification played a key role for lexical processing in the model, we speculate that if the words were written from right to left (e.g. in mW and iW conditions), Mr. Chips should generate more saccades that land on the second half of a word to capture more information about word identity, assuming that the first half of the word were identified beforehand and letter mirroring and inversion do not affect it's lexical access process.

Our hypothesis for eye-movement measures on reading unfamiliar typography are derived from the predicted increase in perceptual difficulty and additional oculomotor demand. Longer fixation durations and shorter saccade amplitudes can be expected for more difficult texts in all four conditions. Increased perceptual difficulty should also result in less skipping cases, but more refixations; both of these predictions are compatible with a reduced average saccade length (Fig.~1b). 

On the level of within-word landing postions, our hypotheses are more specific for the different experimental conditions. Reading words with mirror-inverted letters (mL) or with scrambled letters (sL) will produce within-word landing positions similar to normal reading, since information on letter positions did not change (mL) or did not deviate systematically from normal reading (sL). However, due to inversion of letter positions in reading mirrored-words (mW) and inverted words (iW), we expect readers to shift their eyes further to the right of the word string in the initial saccade and to generate a regressive refixation after that inital saccade (see Fig.~\ref{fig_hypos}a). 

Our study employs a learning paradigm, where participants are required to train reading text in one of the four experimental conditions. The motivation for the learning paradigm is to check the stability of the resulting eye-movement patterns and to exclude the possibility that the results are a short-lived effect due to first exposure to an unfamiliar layout.

In general, we expect that the training will lead to improved performance on the level of global reading measures (average fixation times and mean saccade lengths). Since the scrambled letters (sL) condition represents the least systematic variation of the text layout, we expect that the learning effects are smaller than in the other three experimental conditions. On the level of within-word landing positions, we expect that stable shift of the initial landing positions towards the end of the word strings will be established during learning in the conditions with mirrored words (mW) and inverted words (iW), since the word beginnings are at the end of the manipulated string in these two conditions. For the other two conditions (mL, sL), we expect that within-word landing postions are very similiar to normal reading after training. 

\section{Results}

Reading text with manipulated layout due to mirrored-reversed and inverted letter arrangement produced changes in behavioral measures on a global level (e.g., mean fixation duration, average saccade length) and on the oculomtor level (i.e., within-word fixation locations/saccadic landing positions). While the primary goal of our study was to investigate the possible modulations of oculomotor processes in response to variations of the spatial layout of texts, we start with the presentation of results on global summary statistics to evaluate the overall effects on reading performance as characterized by eye-movement measures. 

\subsection*{Global summary statistics}
This section highlights the significant results for each manipulation across training session in comparison with normal reading. If not otherwise mentioned, results refer to comparisons between the first experimental session (in one of the four conditions mW, iW, mL, or sL) and the control condition (normal reading). 

First of all, global summary statistics from reading normal texts are replicated. In control condition, most of the first-pass fixations move forward to the next neighbouring words (forward saccades \~{41\%}) or skip (skipping saccades \~{26\%}). About \~{22\%} of the fixations move within a word (refixation saccades) and 12\% of them move backward (regression saccades). On average, readers move their eyes 7.82~character spaces forward, 4.23~character spaces backward and fixated on words for 245~ms during reading normal text. Compared to control condition, percentage of forward saccades in experimental condition did not show strong deviation (31\% - 44\%). The percentage of skipping saccades was reduced to less than 5\% in conditions where texts were written against the reading direction (e.g. in mW and iW conditions) and between 10-15\% in other conditions. Interestingly, half of the first-pass fixations observed in conditions where texts were written against reading direction came from refixation saccades, while only a third of first-pass fixations were generated from refixation saccades. Less than 12\% of fixations in reading manipulated text were moving backwards (e.g. regression saccades) with the lowest observed in iW condition. When reading the manipulated texts for the first time, readers fixated on words, on average, about 53-137 ms longer and generated shorter saccades (\~{5-6}~character spaces for progressive saccades and \~{3}~character spaces) than those observed in reading normal text. Note that mean fixation duration in sL condition is about 50~ms longer than those reported in the study by White and colleagues \cite{Whiteetal2008}. Global summary statistics from all sessions are presented in Table \ref{Tab_GlobSum}. 

\begin{figure}[p]
\unitlength1mm
\begin{picture}(150,130)
\put(-2,-5){\includegraphics[width=135mm]{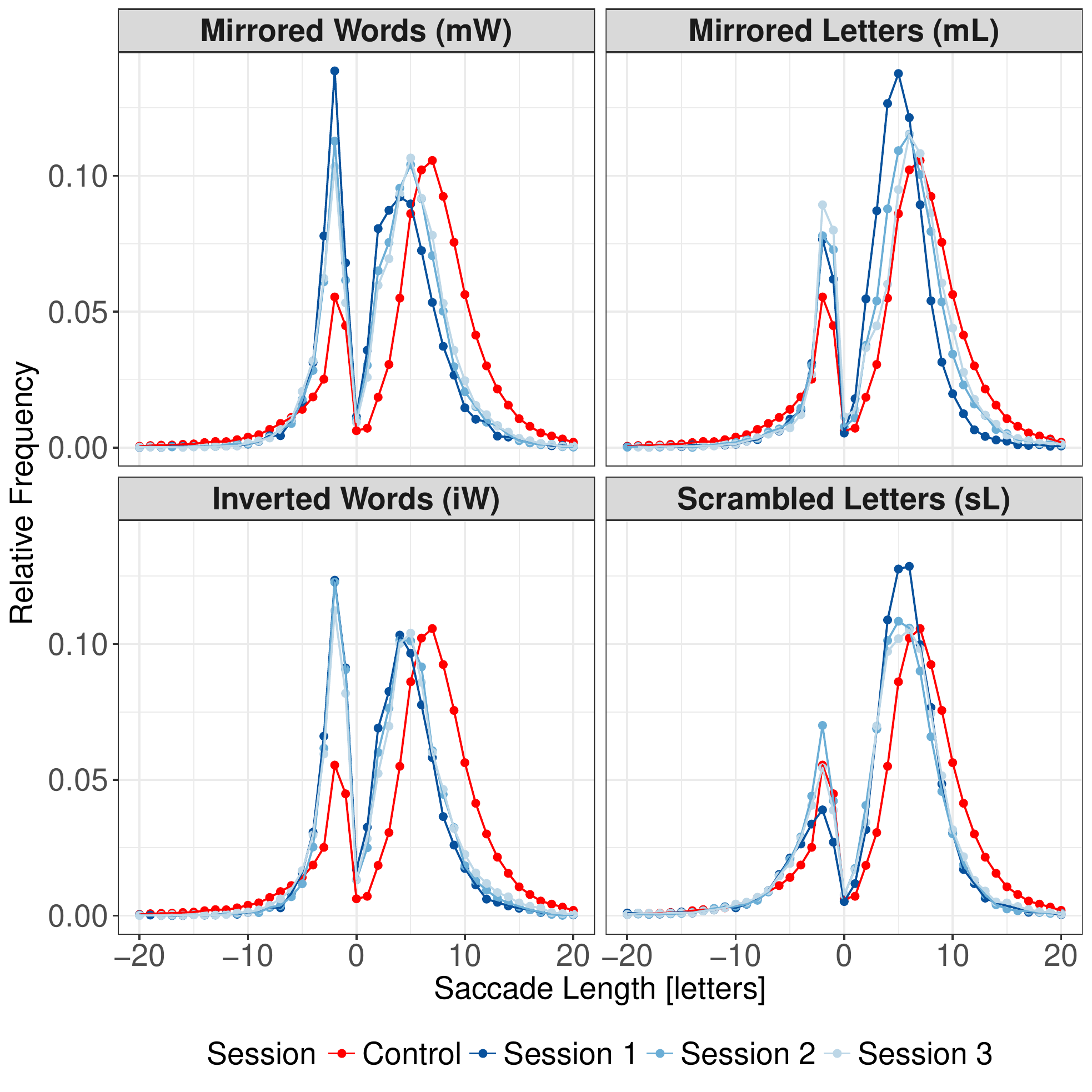}}
\end{picture}
\vspace{5mm}
\caption{\label{fig_sacclen}{}
Distribution of saccade amplitudes across all sessions. Saccades observed in the control condition (normal reading) are marked with red color. Dark blue lines represent data from first session of reading manipulated text. Data from sessions after the two trainings are marked with lighter blue hues. }
\end{figure}

On a qualitative level, we compare the resulting saccade length distributions (Fig.~\ref{fig_sacclen}) with our hypotheses (Fig.~\ref{fig_hypos}b). As expected, the distributions of foward saccades length during reading manipulated text are generally shifted to the left of those from control condition, indicating a qualitatively shorter saccade length generated. Futhermore, more regressive saccades were observed only in reading mirrored-word (mW) and inverted-word (iW) conditions.

\subsubsection*{Single fixation duration and refixation probability}
Results from statistical modeling using linear mixed-effects models for single fixation duration and refixation probability are summarized in Tables \ref{LMM_Sfix}-\ref{LMM_Rfx}. For specification of linear mixed-effects models, see section~\ref{sumstat}.

No significant word length effect was observed in single fixation duration measure for control condition, which in line with the finding reported in study conducted by Kliegl et al. \cite{Kliegletal2004}. Compared to control condition, word-based single fixation durations observed in experimental conditions are significantly longer (all $t$ value $>$ 2). For example, on medium words (5-7 characters), estimated mean single fixation durations in the first experimental sessions were 48-104~ms longer than in control condition (estimated mean: 238~ms). In line with the study by Kowler and Anton \cite{KowlerAnton1987}, texts written against reading direction were more difficult to process. This processing difficulties seemed to be the property of the corresponding type of text manipulation and were more obvious after participants have learned how to read the texts. When readers fixated on medium words only once in the last experimental sessions, the fixation durations in mW (estimated mean: 309~ms) and and sL (estimated mean: 326~ms) conditions were shorter compared to those observed in mW and iW conditions with estimated mean of 570~ms and 455~ms respectively. 

Similar to single fixation measures, there was no significant word length effect observed in first of multiple fixation duration measures for control condition. The mean of first fixation duration in sL condition did differ significantly across all sessions compared to those observed in control condition. In mL condition, the duration of first fixation measures was significantly longer when readers read the manipulated text for the first time, but did not differ significantly after trainings. Interestingly, the mean of first fixation duration was significantly longer in mW and iW conditions compared to control condition. For example, on medium words, estimated mean durations of the first fixation in both conditions were about 90~ms longer than those estimated for control condition. The difference remained significant even after trainings (estimated means session 3: $mW=465.32$; $iW=414.38$).    

On refixation probability measure, we replicated the word length effect in both control and experimental conditions (all $p < 0.01$). Long words were more likely to be refixated than short words. However, when the texts were manipulated, readers were more likely to refixated the words. For example, the estimated refixation probability on medium words in control condition was 0.16. However, readers in experimental conditions were twice more likely to fixate the words with the same lengths (estimated RFP: $mL=0.42$; $sL=0.65$; $mW=0.44$; $iW=0.56$). Interestingly, after training to read texts with words written against the reading direction (e.g. mW and iW conditions), readers were almost always fixated on medium and long words (estimated RFP above 0.87 on the third session). 

Do these manipulation types have something in common or do they generate different effect on fixation duration and probability measures presented above? The results from separate models which estimated the different effect size of manipulation types (see Table \ref{LMM_effect}) confirmed that some manipulation type generated greater effect than the others. All of the three models showed similar trends. The effect in mL condition was significantly different from control condition, but adding the effect from sL conditon to the mean of the two conditions (control and mL conditions) did not yield a significant gain on effect size. However, adding the effect from mW condition and iW conditions yielded significant gains on the effect size. The results from various measures showed that of all manipulation types tested in the current study, the iW condition generated the largest effect. 

\begin{figure}[p]
\unitlength1mm
\begin{picture}(150,140)
\put(0,0){\includegraphics[width=140mm]{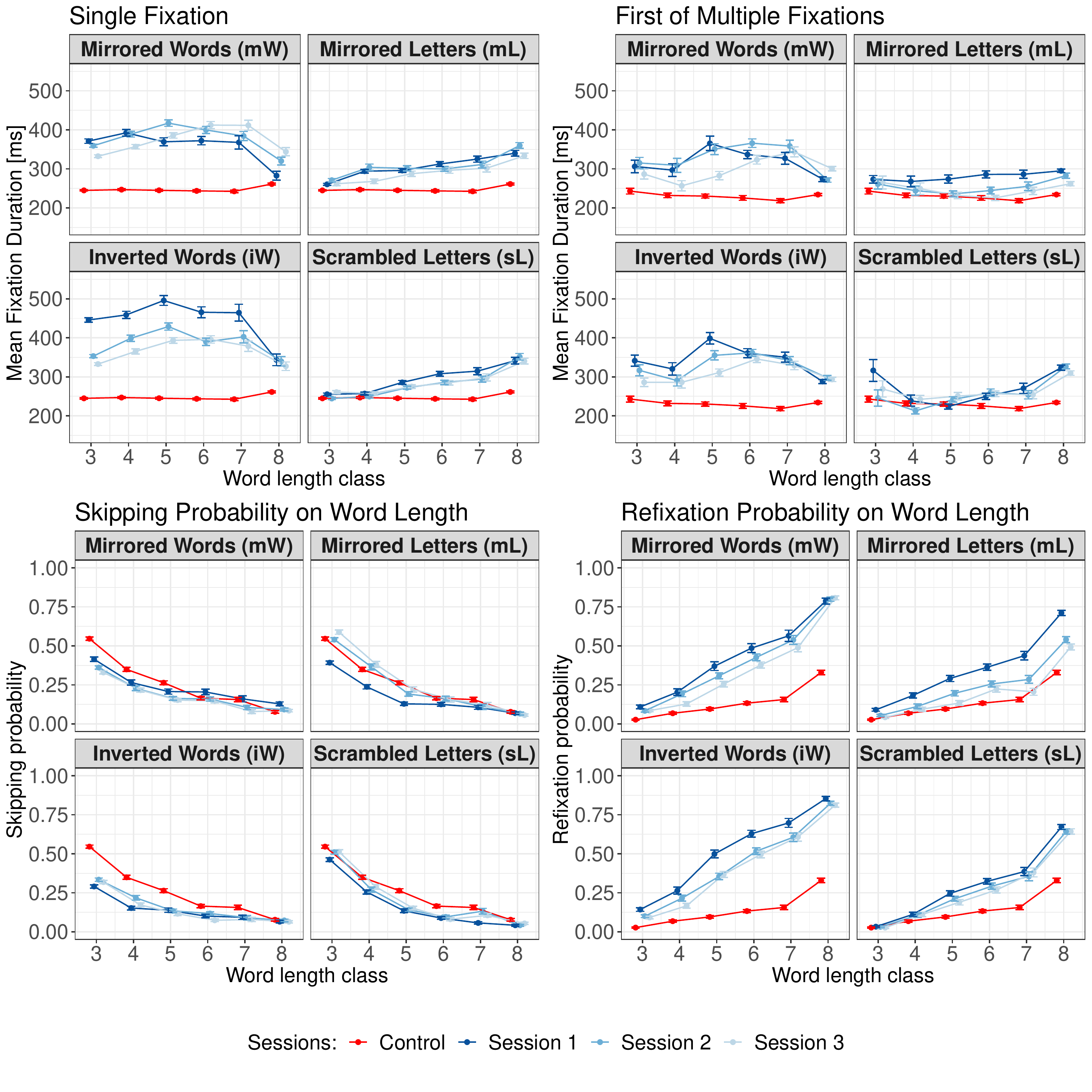}}
\end{picture}
\vspace{-8mm}
\caption{\label{fig_dursprob}
\textit{Upper row.} Mean fixation duration of single fixation (left) and first of multiple fixations as the function of word length classes. \textit{Lower row.} Skipping (left) and refixation probabilities as the function of word length classes. Word with length less than 3 characters are grouped in word length class 3; long words ($> 8 characters)$ are grouped in word length class 8. Error bars represent the standard error of the means. Red line and dots represent data from reading normal text. The dark blue color represents data from the first experimental session. The lighter blue hues indicate data from the last two experimental sessions. }
\end{figure}

The effects of word lengths on duration of single fixation and first of multiple fixations, as well as on skipping and refixation probabilites are visualized in Fig.~\ref{fig_dursprob}.  

\subsection*{Landing positions distributions and launch-site effect}
A first glance at the resulting distributions for within-word landing positions indicates differences between normal reading and manipulated texts (Appendix \ref{landpos_iW}-\ref{landpos_sL}). Except for short words (i.e., word length up to 4 chars), we found increased leftward shifts across all experimental conditions, where with mirrored words (mW) and inverted words (iW) showed stronger shifts compared to the other two manipulations (mL, sL). Since shorter saccades were generated in experimental conditions, we analyses only saccades launched up to 5 character away from the word beginning, resulting in more overshoots. Undershoots were typically observed on longer saccades launched furhter than 7 characters away and more prominent on long words \citep{McConkieetal1988,Nuthmannetal2005,KruegelEngbert2010}. Based on Bayesian fits of the distribution (see Methods), we obtained mean landing positions that were used for further analyses.

\begin{figure}[p]
\unitlength1mm
\begin{picture}(150,100)
\put(0,0){\includegraphics[width=140mm]{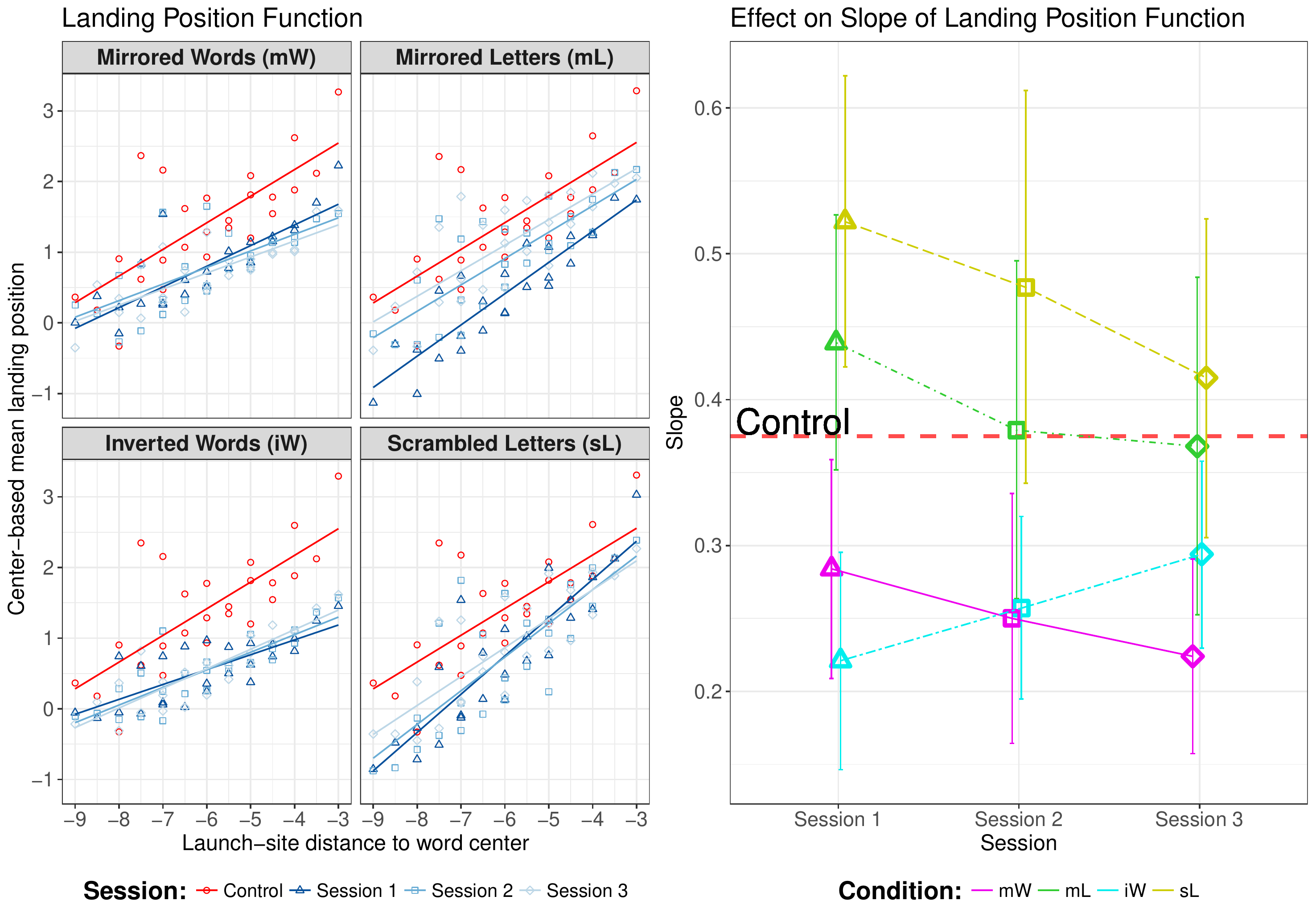}}
\put(0,-3){(a)}
\put(86,-3){(b)}
\end{picture}
\vspace{2mm}
\caption{\label{fig_mcconkie}
Analysis of the launch-site effect. (a) Landing position function: Mean center-based landing position as a function of center-based launch-site distance for all experimental conditions. Red line and dots represent data from normal reading session. Data from the first experimental sessions are presented in dark blue color. The lighter blue hues represent the last two experimental sessions.  (b) The estimated slope parameter $\lambda$ (and the 95~\% confidence interval)
for all experimental conditions as a function of the experimental session. The red horizontal dashed line represents the estimated slope for reading normal text (baseline). Results for reading mirrored words (mW) and inverted words (iW) are presented with magenta and cyan lines, resp. Green and yellow lines represent estimations for reading texts with mirrored letters (mL) and scrambled letters (sL).}
\end{figure}

In Figure \ref{fig_mcconkie}a, center-based mean landing sites are plotted against center-based launch sites for forward saccades and different lengths of the target words. The slope parameters from landing position functions are plotted for all sessions in Figure \ref{fig_mcconkie}b. The estimated slope of ($\lambda_0=0.37$) was observed in reading normal text. Compared to the value from control condition, both conditions iW ($\lambda_1=0.31$,$\lambda_2=0.25$,$\lambda_3=0.23$) and mW ($\lambda_1=0.24$, $\lambda_2=0.26$, $\lambda_3=0.29$) generated shallow slopes, which indicates a tendency to reduced oculomotor control. In contrast, conditions mL ($\lambda_1=0.44$, $\lambda_2=0.38$, $\lambda=0.37$) and sL ($\lambda_1=0.52$, $\lambda_2=0.48$, $\lambda=0.41$) generated greater slope values ($\lambda$) compared to reading normal text, indicating a tendency to increased oculomotor control. Interestingly, except in the mW condition, the $\lambda$ values approach the value of normal reading after trainings.

In the Bayesian model, the slope parameter $\lambda$ represents the strength of oculomotor control {\em and} the relation between observational error of the target location ($\sigma_0^2$) and standard deviation of prior distribution ($\sigma_T^2$). In extreme cases, a value of slope near to 0, where  $\sigma_0^2$ $\,\to\,$0, the eyes always land on the target (i.e., the word center) regardless of where the saccade started. In the other extreme case, the a slope value of $\lambda\to 1$ indicates the absence of a target selection process in saccade planning, so that the eyes generate random constant saccade lengths (from a uniform distribution). Therefore, our data show that saccades from reading text composed of words with reversed letter sequences (i.e., mW and iW) tend to land precisely on the target location (word center) on average, while reading manipulated text with normal letter order (i.e., mL and sL) tend to generate saccades similar length.  

Comparing the results with our hypotheses, there is no dramatic effect of a shift of the mean landing position towards the word ends. Therefore, we ran a post-hoc analyses for single and two-fixation cases in the next section.

\subsection*{Landing positions for single vs.~two-fixation cases}
{\sl Single fixation cases.}
Only cases where exactly one fixation land on a word were considered. Relative frequency of fixation landing position were calculated based on word length. To obtain estimates for the mean $\mu_{SF}$ and standard deviation $\sigma_{SF}$ for the landing position distribution of each word length, a grid search method (in steps of 0.01) with a minimum-$\chi^2$ criterion was applied. Analysis of variance was conducted to statistically compare the manipulation effects on mean ($\mu_{SF}$) and standard deviation ($\sigma_{SF}$) of landing position distributions. 

No significant difference was observed for the mean ($\mu_{SF}$) of the landing position distribution of single fixation cases between control and experimental conditions ($F[1,30]=2.364$, $p=0.135$). The main effect on standard deviation ($\sigma_{SF}$) of within-word landing position distribution of single fixation cases yielded an $F$-ratio of $F[1,30]=32.7$,$p<0.000$, indicating a significant difference between reading normal text (M=$2.95$,SD=$0.84$) and manipulated text (M=$1.51$,SD=$0.55$). When words were fixated only once, readers' mean landing position did not change across the manipulation types. However, the precision of landing on selected target increased as the text display deviated from normal presentation. 
 
\begin{figure}[p]
\unitlength1mm
\begin{picture}(150,160)
\put(-8,0){\includegraphics[width=155mm]{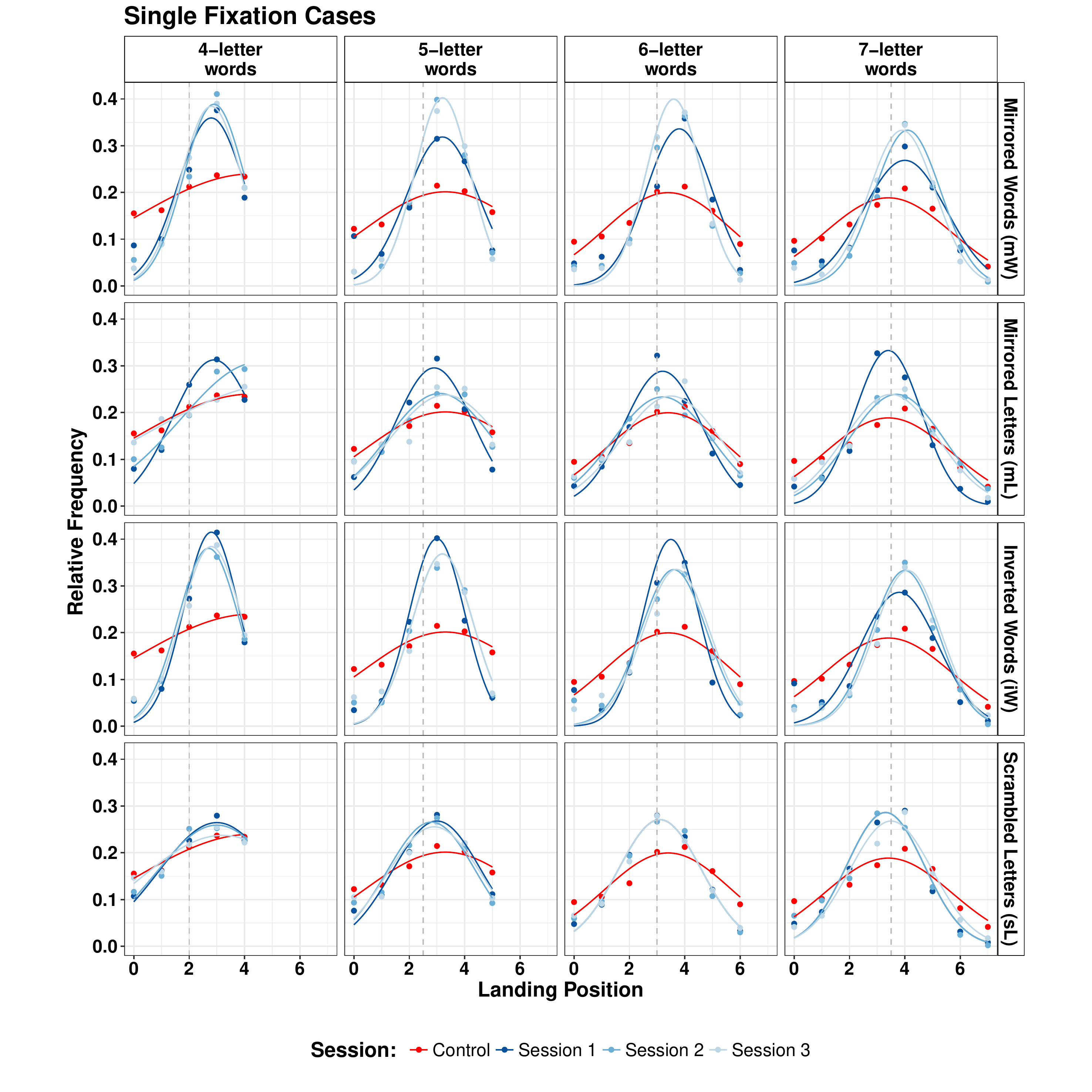}}
\end{picture}
\vspace{-1mm}
\caption{\label{fig_singlefix}
Within-word landing position distributions for single-fixation cases. Columns relate to different word lenghts. Rows refer to experimental manipulations. Different colors indicate control condition (red) and experimental conditions (blue hues).}
\end{figure}

Among our key results is that the two conditions where the word beginnings were located at the end of the manipulated word strings (i.e., conditions mW and iW) produced only a slight shift of the average landing position. This finding turned out to be robust and remained observable even after trainings.  Figure \ref{fig_singlefix} presents data for single fixations. The finding did not support our hypothesis that readers targeted the second half of the word strings during reading mirrored-word (mW) and inverted words (iW). On a qualitative level, there is little adaptivity of the oculomotor system.

\begin{figure}[p]
\unitlength1mm
\begin{picture}(150,100)
\put(0,-2){\includegraphics[width=140mm]{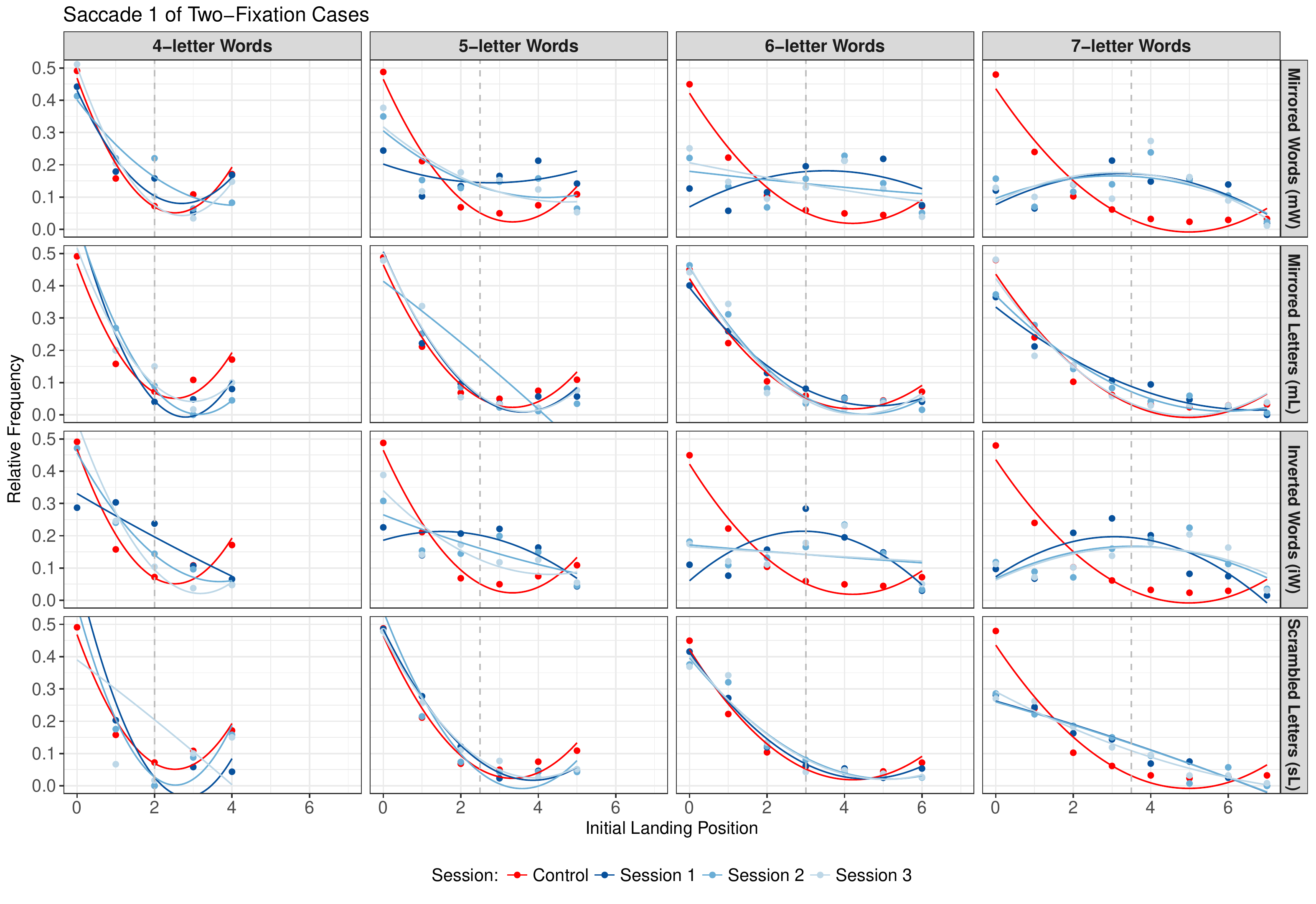}}
\end{picture}
\vspace{-1mm}
\caption{\label{fig_twofix}
Within-word landing position distribution for the initial landing position (first saccade) in two-fixation cases.}
\end{figure}

Since in our hypothesis, an initial saccade towards the second half of the word would require an additional refixation in the first half of the word, we ran a post-hoc analysis of all two-fixation cases. The corresponding plot in Figure \ref{fig_twofix} presents data for the initial landing position (first saccade into the word) from all cases, where the word was fixated exactly two times. In contrast to typical OVP effect on refixation probability, plotting the initial of two fixations give us a better understanding of word targeting in saccade planning. 

{\sl Two-fixation cases.}
For this analysis, we considered cases when exactly two fixations on a word were observed. The landing position distribution of the initial fixation was fitted to a quadratic polynomial, i.e.,
\begin{equation}
\label{eq_firstoftwofix_landpos}
y = A + B\cdot L(x-C)^2 \;,
\end{equation}		
where $x$ denotes the initial landing position and \textit{y} is the relative frequency of the fixation. The parameter \textit{A} represents the actual relative frequency of the initial landing position. The parameter \textit{B} is the slope of the parabolic curve, representing the within-word maximum or minimum relative frequency of the landing position. The parameter \textit{C} reflects within-word position where the relative frequency was at the minimum or maximum, depending on the value of parameter \textit{B}. In general, the value of the parameter \textit{B} is assumed to be positive, resulting in a distribution of landing position qualitatively similar to the optimal viewing position (OVP) curve but with different interpretation: when a word is fixated exactly twice, the initial fixations tend to on word edges more often than on the word center. The estimation of the three parameters representing the characteristics of the landing position of the first of two fixation cases was conducted based on a maximum likelihood method using the bbmle package \citep{bbmle2017} for R studio. The estimated parameters are summarized in Table~\ref{Quad_N}-\ref{Quad_sL}.

In two-fixation cases under normal reading conditions, readers tend to initially land their eyes close to the word beginning with a subsequent secondary fixation further into the word (most refixations are forward directed). Similar patterns were observed in the condition where manipulated texts were written in typical direction from right to left (mL and sL). However, what is most remarkable for our study, in the two conditions (mW and iW) with reversed letter sequences, the initial landing positions deviated from those of normal reading (for detailed estimated parameters, see Table~\ref{Quad_mW} and \ref{Quad_iW}). When a word was fixated more than once, the first saccade landed mostly behind the location of the word center in the second half of the word strings; thus, there is clear top-down adaptivity in the two conditions where letter sequence is reversed from right to left. The effect was stronger for longer words (see Fig.~\ref{fig_twofix}). 
   
These findings support our hypothesis that readers first target the right part of the word strings and then moved the eyes backward, following the writing direction.
Topical subheadings are allowed.

\section{Discussion}

Natural text reading requires efficient coordination of cognitive and oculomotor control processes. Cognitive principles are essential for the selection of an upcoming target word, however, it is the oculomotor system that provides the machinery to move the gaze to the region of the identified target word. In the current study, we set out to investigate whether ongoing cognition is able to overwrite default oculomotor control when the reader is confronted with manipulated (mirrored-reversed, inverted, and scambled) text layouts. 
 
On a global level, changing the display and positions of letters in a word modulates eye-movement statistics. Since mirrored-reversed and other manipulated words are more difficult to process, a general tendency of longer average fixation durations and shorter mean saccade amplitudes is observed. These findings are in line with previous studies \citep{KowlerAnton1987,Rayner2006}. Interestingly, similar with findings from Kowler and Anton \cite{KowlerAnton1987} and Kolers and Perkins \cite{KolersPerkins1975}, orthographical manipulations from our current study can be grouped into two categories based on their writing directions. Difficulties increased dramatically, when readers read texts written against the reading direction as in mirrored words (mW) and inverted words (iW) conditions, where the letter-sequence of words is written from right to left within the displayed string due to mirroring or inversion. Our interpretation is supported by the fact that the observed deviations remained stable even after trainings. In contrast, during the manipulations that kept the letter-sequence in the reading direction (i.e., from left to right) in mirrored-letter word (mL), the patterns observed in eye movement measures (e.g., word-length effect) remained similar to those of normal reading and, after training, approached the behavior observed in normal reading. Finally, in our fourth experimental condition, we analyzed reading text composed of scrambled-letter (sL) words. Since letters in a word were randomly scrambled, there was no systematic way that the oculomotor behavior could adapt to process the new text layout. 

One of our motivations to run the current study was to obtain a detailed picture on within-word landing positions as the most important signature for oculomotor control. Mean landing position distribution of single fixations in reading manipulated texts does not significantly differ from reading normal text, although there was a slight shift observed in mirrored words (mW) and inverted words (iW) conditions. However, readers tend to increase their precision in landing on word center when texts presentation was manipulated. The finding supports the idea that word center serves as saccade target location \citep{EngbertKruegel2010}. Analysis of mean landing position of forward saccades based on different launch sites demonstrated a general left shifts in experimental conditions compared to those observed in control condition. Especially in conditions where words are written against reading direction such as in mW and iW conditions, there is no clear preference for the eyes to initially land in the second half of the word strings (where the beginning of a word is located in this manipulation). This is contrary to our hypothesis. Interestingly, the peaks of the landing position in all manipulation types were shifted toward word centers, not necessaryly toward the beginning of word strings, compared to the peaks observed in reading normal text. In some rare undershoot cases in our data (e.g. launch site: -4, word length: 7 in iW and mW conditions), the peaks in experimental condition shift rightward of the peak observed in reading normal text. Furthermore, the variance of the initial landing position distribution was much smaller, meaning that the eyes landed more precisely on the word center, which served as the theoretical target position. Given the precision of the landing position distributions and the highest likelihood to land on word center, we speculate that the leftward shifts were actually moving toward the word centers, not word beginning. Regarding the general improved precision of landing position distributions, a possible explanation is that with longer average fixation durations under our manipulations, the oculomotor system had more time to prepare the next saccade, which could result in a reduced saccadic error \citep{McConkieetal1988}. An alternative explanation is that unusual presentation of text was more salient and popped up parafoveally to enable precise saccadic targeting \citep{Hyona1995}. 

Additionally, modulation of manipulation types on launch-site effect was observed. In conditions where texts were written against reading direction (from right to left) we observed a reduced launch-site effect, while an increased launch-site effect was observed in conditions where texts were written normally. However, we are not sure if the difference effects are caused by the change in reading direction or increased processing difficulties. Nevertheless, the finding could also be explained in the theoretical framework of a Bayesian model of sensorimotor integration that we applied to reading \citep{EngbertKruegel2010,KruegelEngbert2014}. According to Bayes' rule, the observed landing position distribution is the posterior distribution in a sensorimotor transform based on prior knowledge of typical target positions. If we assume that the prior knowledge is uninformative under unfamiliar text layout, then the posterior distribution would shift toward the likelihood distribution, which is assumed to be unbiased with respect to the target word center. Therefore, in the Bayesian model, we could have expected a more centered landing-position distribution if we assumed that the experimental manipulations induce less prior knowledge on the possible target positions. Our finding demonstrated that by simply changing the letter-sequence information of words, we could obtain estimations for the slope parameter $\lambda$ other than the typical value of 0.5. As a consequence, reading models development should aim for integrating a process-oriented model in generating saccade lengths.

The most striking finding in the current study is the effect of text-display manipulation on initial landing position of two-fixation cases. When a word shows reversed letter ordering (hence a change in their spatial information) as a consequence of the experimental manipulation, we observed a clear shift into the second half of the word string when two fixations were generated. This finding demonstrates that the oculomotor system is able to adapt to display changes. Given that our hypothesis requires more than two fixations, analysing the initial landing positions in refixation cases is more reasonable to test the hypothesis. 

Furthermore, our study also demonstrates that the typical usage of mean fixation position as a dependent variable to characterize oculomotor control is not specific enough to characterize control processes underlying reading. Since reading is a complex process, data delivered are usually complicated and require various analysis methods and inferences. Further systematic analyses and advanced mathematical modeling are required to investigate the dynamical processes underlying principles of oculomotor control and their interaction with ongoing cognition during reading. 

\section{Methods}

\subsection*{Participants}  
A group of 37 participants (27 females, 10 males), aged between 16 and 39 years, received a total of EUR~70 for taking part in four 45-minute lab sessions and six 30-minute training sessions. They were all naive with respect to the purpose of the experiment. Participants reported normal or corrected-to-normal vision and declared their informed consent, p. The experiment conformed to the Declaration of Helsinki. Informed consent was obtained for experimentation by all participants. According to the standards of Deutsche Forschungsgemeinschaft (German Research Foundation) and German Society for Psychological Research, ethics committee approval was not required for this study.

\subsection*{Apparatus, Material \& Procedure}
Participants were assigned to four different groups based on four types of text manipulation, namely mirrored-word text (mW), mirrored-letter text (mL), inverted-word text (iW), and scrambled-word text (sL). Each participant did a total of four lab sessions and six training sessions at home via web-based interface. 

For lab sessions, participants read an excerpt from the German version of the novel “The Adventure of the Empty House” \citep{EmptyHouse2009}. They were seated at a viewing distance of 70cm in front of a 19-inch Mitsubishi Diamond Pro 2070 Monitor (screen resolution 1,280$\times$1,024 pixels, refresh rate 100~Hz) with the head supported by a chin rest. The stimuli (Courier font, size 18, black) were presented on vertical center line of the computer display with gray background color. Eye movements from both eyes were recorded using an EyeLink 1000 System (SR Research, Osgoode/Ontario, Canada) with a sampling rate of 1000~Hz and spatial resolution better than 0.01$^\circ$. At the end of the session, participants had to answer three questions related to the text they have just read. A maximum of 600 lines of text were presented over all lab sessions.

For training sessions, participants read an excerpt from the novel ``Small World'' \citep{Suter97} by visiting a website created for the experiment\footnote{Using ShinyApps by RStudio accessible via {\tt http://www.shinyapps.io}}. After logging in, participants could read the manipulated text, which was presented as a line of max.~85 characters at the center of the screen, without time limit. When finished, they could go to the next line by clicking the right arrow or return to the previous page by clicking the left arrow. One training session should last at least 30 minutes. After logging out, participants received three questions via E-mail, which should be answered as soon as possible. No limit of the number of lines presented in training sessions was enforced.

The complete procedures of the experiment went as follows: During the first lab session, participants read normal text (a total of 150 lines of maximum 85 characters). After a two-hour break, participants read manipulated text in the second session, which lasted up to 45 minutes or when 150 lines were read. At home, participants were required to read manipulated text for two 30-minute sessions on the website before the third lab session. At the third lab session, participants continue reading manipulated text from where they left off at the previous session. Participants should conduct four 30-minute training sessions before taking part in the last lab session.

\subsection*{Data Preparation}
Data containing blinks were discarded from the analysis. Saccades and fixations were detected using a velocity-based algorithm developed by Engbert et al. \cite{microsacTB}. As a result, a total of $380,292$ fixations were detected. From this data set we excluded fixations based on the following criteria (i) fixations on the first and last words of a sentence as well as the first and the last of participant's fixation sequence, (ii) fixations with duration less than 20~ms or longer than 1000~ms, fixations landing outside the text rectangle, and saccades shorter than one character space (12~pixels) or longer than 25~character spaces were removed from the analysis. Trials containing fixation duration longer than 2000~ms and less than three fixation points after filtering were excluded from analyis. The remaining $236,937$ fixations are the valid fixations (see Table~\ref{Tab_GlobSum}).

\subsubsection*{Summary statistics} \label{sumstat}
In a global analysis, we computed statistics of fixation durations and saccade lengths of all valid fixations. For local processing, fixation duration and probability were grouped based on word length. Word lengths were grouped into three classes: short ($\leq$~4 characters), medium (5-7 characters) and long ($>$~7 characters) for further statistical analysis.

{\sl Fixation durations.} 
Visual information processing in reading is marked by the time spent on fixating words. Statistical analysis on single fixation For complimentary analysis on possible differences in processing All cases in which a word received exactly one fixation generates single fixation duration (SFD). In the cases in which a word received more than one fixations, the duration of the first and the second of those fixations were calculated for evaluating the first of multiple fixations (FMD) and second of multiple fixations (SMD). All of those fixation duration measures were conducted on the first-pass reading data set. The sum of all fixations on a word, regardless the saccadic types, generated the total viewing time (TVT). For further statistical analysis, the fixation duration measures on each word were transformed into their logarithmic values.

{\sl Fixation probability.}
As measures of fixation probability we computed skipping probability (SKP), single fixation probability (SFP), and refixation probability (RFP) on first-pass reading data. Additionally, we calculated regression probability (RGP). For further statistical analysis, each word was assigned a logical value of 1 if it is skipped (SKP), fixated only once (SFP), fixated more than once (RFP) or a target of regressive saccades (RGP). Otherwise, a logical value of 0 was assigned to the word.

To capture the changes of the manipulation effects on word processing difficulties across sessions, linear-mixed model analysis (3x4 factors) was conducted on the dependent variables single fixation duration (SFD) and refixation probability (RFP) as a function fixed word lenght effect and random effects generated by participants and sentences. The model was contrasted against the word length effect during normal reading ({\tt sess = 0}). We used the lme4 package for R-Language of Statistical Computing \citep{BatesLME4} for estimating fixed (word length classes) and random (participant and sentence) coefficients. Specifically, each fixation duration and probability measure is modeled as a function of fixed word length effect ({\tt WLC} across sessions ({\tt sess}), with a fully parameterized variance-covariance matrix for participant ($1+\mbox{\tt WLC}+\mbox{\tt sess}|\mbox{\tt pID}$) assuming participants may generate different slopes for each word length and sentence/line number $(1|\mbox{\tt sID})$ with the assumption that the intercept for each sentence do not vary accross sessions and word lengths. These model used the default treatment contrast. For example, the log-transformed measure of single fixation duration ({\tt SFD}) is modeled as 
\begin{equation}
\label{eq_LME_SFD}
\mbox{\tt lmer}(\log(\mbox{\tt SFD}) = \mbox{\tt WLC}*\mbox{\tt sess} + (1+\mbox{\tt WLC}+\mbox{\tt sess}|\mbox{\tt pID}) + (1|\mbox{\tt sID})),
\end{equation}
while refixation probability ({\tt RFP}) are modeled as
\begin{equation}
 \label{eq_LME_skip}
\mbox{\tt glmer}(\mbox{\tt RFP} = \mbox{\tt WLC}*\mbox{\tt sess} + (1+\mbox{\tt WLC}+\mbox{\tt sess}|\mbox{\tt pID}) + (1|\mbox{\tt sID})),
\end{equation}

because the \mbox{\tt glmer()} function allows the statistical analysis of binary outcome. The models were estimated based on restricted maximum likelihood (REML). Furthermore, a 3x5-factor model using helmert constrast was conducted separately for each dependent measure with word lengths and manipulation types ({\tt type}) as fixed main effects to investigate the different effect sizes among different manipulation types. The predictor {\tt type} for reading normal text was used as baseline ({\tt type = 0}). 

\subsubsection*{Initial Landing Position} 
Within-word landing-position distributions in reading are broad and overlap with neighboring words. As a consequence, observations of word-based landing positions in a reading experiment are truncated at word boundaries \citep{McConkieetal1988,EngbertNuthmann2008}. In order to derive estimates of the means and standard deviations of the landing-position distribution, truncated normal curves for the distributions of within-word fixation positions were fitted using Bayesian parameter inference \citep{Kruschke2014} provided by the package \textit{rjags} \citep{rjags2016} in the R environment.

Here we considered only cases where readers made forward saccades. Fixation data for each experimental condition and each experimental session were grouped based on word length and launch-site distance leading to 16 different word-length and launch-site specific data subsamples per condition and reading session. For each data subset $S_i$, we estimated a two-dimensional posterior distribution over the parameters mean $\mu$ and standard deviation $\sigma$ of the underlying Gaussian landing-position distribution according to 
\begin{equation}
p(\mu,1/\sigma^2|S_i)=\frac{p(S_i|\mu,1/\sigma^2)p(\mu,1/\sigma^2)}{\int\!\int p(S_i|\mu,1/\sigma^2)p(\mu,1/\sigma^2) \, \mathrm{d}\mu \mathrm{d}1/\sigma^2} \;.
\end{equation}
We assumed that observations of landing positions are generated by a normal-density likelihood function $p(S_i|\mu,1/\sigma^2)$ with mean $\mu$ and precision $\tau=1/\sigma^2$. Furthermore, with $p(\mu,1/\sigma^2)$ we specified a normally distributed prior on the mean $\mu$ with mean M and precision $T$ and a prior on $\tau$ distributed as a gamma density distribution with shape parameter $A$ and rate $B$ (see \cite{Kruschke2014}). The parameters M, T and A, B for the prior over $\mu$ and $\tau$, resp., were derived from the data in the following steps: For the control condition, landing-position distributions for each word length and launch-site distance were independently fitted by a truncated Gaussian function and the parameters of these fits were used to estimate the parameters of the empirical prior distribution. For reading manipulated text we also used the parameters of the prior of the control condition for estimating landing distributions of the first reading session and updated the prior for the later sessions systematically based on the results of the previous reading session.

\section{References}
\renewcommand{\bibsection}{}
\bibliography{Library}

\section{Acknowledgements}
This work was supported by Deutsche Forschungsgemeinschaft (grant EN 471/15-1 to R.~E.). 

\section{Author contributions statement}
R.E. and A.K. developed the study concept and the study design. Testing and data collection were performed by J.C..  J.C. and A.K. performed the data analysis and all authors contributed to the interpretation. J.C. and A.K. drafted the manuscript, and R.E. provided critical revisions. All authors approved the final version of the manuscript for submission.

\section{Additional information}
\textbf{Competing interests} 
The authors declare no competing interests.

\begin{appendix}
\section{Appendix}

\setcounter{table}{0}
\renewcommand{\thetable}{A\arabic{table}}
\setcounter{figure}{0}
\renewcommand{\thefigure}{A\arabic{figure}}

\begin{landscape}
\begin{table}[p] 
\footnotesize
\caption{Global summary statistics.}
\label{Tab_GlobSum}
\begin{tabular}{|lc|c|ccc|ccc|ccc|ccc|}
\hline
\multicolumn{2}{|l|}{Condition} & Control & \multicolumn{3}{c|}{Mirrored Words (mW)} & \multicolumn{3}{c|}{Mirrored Letters (mL)} & \multicolumn{3}{c|}{Inverted Words (iW)} & \multicolumn{3}{c|}{Scrambled Letters (sL)} \\
\hline\hline
Session  	& 	 & 0 	 & 1 	 & 2 	 & 3 	 & 1 	 & 2 	 & 3 	 & 1 	 & 2 	 & 3 	 & 1 	 & 2 	 & 3		\\
\hline
Detected  	& N  & 82487 & 21598 & 27500 & 26957 & 26677 & 24205 & 22246 & 22531 & 24104 & 25728 & 25549 & 26648 & 24062  	\\
fixations 	&    &       &       &       &       &       &       &       &       &       &       &       &       &      	\\
\hline
Discarded 	& N  & 40377 & 10351 & 12251 & 11691 & 12143 & 11639 & 11271 & 10340 & 10766 & 10933 & 11361 & 11350 & 10597 \\
fixation  	& \% & 48.95 & 47.93 & 44.55 & 43.37 & 45.52 & 48.09 & 50.67 & 45.89 & 44.66 & 42.49 & 44.47 & 42.59 & 44.04 \\
\hline
Valid  		& N  & 42110 & 11247 & 15249 & 15266 & 14534 & 12566 & 10975 & 12191 & 13338 & 14795 & 14188 & 15298 & 13465 \\
fixation 	& \% & 51.05 & 52.07 & 55.45 & 56.63 & 54.48 & 51.91 & 49.33 & 54.11 & 55.34 & 57.51 & 55.53 & 57.41 & 55.96 \\
\hline
First-pass  & N  & 32343 & 6925  & 11166 & 11620 & 11200 & 10061 & 8809  & 9479  & 10428 & 11912 & 10269 & 10034 & 8836  \\
fixation  	& \% & 76.81 & 61.57 & 73.22 & 76.12 & 77.06 & 80.07 & 80.26 & 77.75 & 78.18 & 80.51 & 72.38 & 65.59 & 65.62 \\                                     
 \hline                                   
Forward   	& N  & 14956 & 2960  & 5386  & 5725  & 5636  & 4379  & 3909  & 4050  & 4653  & 5363  & 4887  & 4219  & 3843  \\
saccades   	& \% & 46.24 & 42.74 & 48.24 & 49.27 & 50.32 & 43.52 & 44.38 & 42.73 & 44.62 & 45.02 & 47.59 & 42.05 & 43.49 \\
\hline
Skipping   	& N  & 9963  & 516   & 834   & 931   & 1476  & 2193  & 1964  & 431   & 674   & 774   & 1973  & 1999  & 1807  \\
saccades   	& \% & 30.8  & 7.45  & 7.47  & 8.01  & 13.18 & 21.8  & 22.3  & 4.55  & 6.46  & 6.5   & 19.21 & 19.92 & 20.45 \\
\hline
Refixations & N  & 7357  & 3436  & 4933  & 4957  & 4082  & 3486  & 2930  & 4989  & 5096  & 5772  & 3400  & 3805  & 3170  \\
 			& \% & 22.75 & 49.62 & 44.18 & 42.66 & 36.45 & 34.65 & 33.26 & 52.63 & 48.87 & 48.46 & 33.11 & 37.92 & 35.88 \\
\hline
Regressions & N  & 3692  & 693   & 808   & 854   & 734   & 695   & 638   & 549   & 584   & 649   & 992   & 985   & 966   \\
 			& \% & 11.42 & 10.01 & 7.24  & 7.35  & 6.55  & 6.91  & 7.24  & 5.79  & 5.6   & 5.45  & 9.66  & 9.82  & 10.93 \\
\hline    
fixation 	& mean & 245 & 354 & 351 & 337 & 299 & 277   & 266   & 383   & 344   & 329   & 306   & 300  & 304  \\
duration 	& [ms] &     &     &     &     &     &       &       &       &       &       &       &      &      \\
\hline
forward  	& mean & 7.82& 5.19& 5.58& 5.81 & 5.53 & 6.56  & 6.86 & 5.25  & 5.55  & 5.81 & 6.18  & 6.15  & 6.38  \\
saccade  	&[char]&     &     &     &      &      &       &       &       &       &       &       &      &      \\
amplitude	&      &     &     &     &      &      &       &       &       &       &       &       &      &      \\
backward  	&mean & 4.23 & 2.82& 2.97 & 3.03 & 3.01 & 2.88 & 2.67  & 2.63  & 2.57  & 2.79  & 4.66  & 3.95 & 4.25 \\
saccade  	&[char]&     &     &      &      &      &      &      &        &       &      &      &      &    		\\
amplitude 	& 	   &     &     &      &      &      &      &      &        &       &      &      &      &          \\
\hline

\end{tabular}
\end{table}  
\end{landscape}

\newpage 

\begin{table}[h]
\footnotesize
\caption{Results from Linear Mixed-Effects: Single Fixation Duration}
\label{LMM_Sfix}
\begin{tabular}{|l|l|c|c|c|}
\hline \hline
Condition 				&              	& \textit{b}   & \textit{SE}  & \textit{t value}      \\
\hline
Mirrored Letters (mL)  	& Intercept  	& 5.47  & 0.04 & 124.53  \\
		 				& WLC          	& 0.02 	& 0.02 & \textbf{1.15}  \\
 						& Session 1    	& 0.18  & 0.02 & 9.69  \\
 						& Session 2    	& 0.10  & 0.02 & 5.32 \\
 						& Session 3    	& 0.09  & 0.02 & 4.22  \\
 						& WLC:Session 1	& 0.06  & 0.01 & 7.79  \\
  						& WLC:Session 2	& 0.07  & 0.01 & 9.38  \\
 						& WLC:Session 3	& 0.06  & 0.01 & 8.30  \\
\hline
Mirrored Words (mW)  	& Intercept 	& 5.44  & 0.04 & 142.66 \\
						& WLC       	& 0.02  & 0.02 & \textbf{0.87}  \\
						& Session 1     & 0.31  & 0.07 & 4.54  \\
 						& Session 2     & 0.31  & 0.06 & 5.11  \\
 						& Session 3     & 0.29  & 0.06 & 5.10  \\
  						& WLC:Session 1 & -0.14 & 0.01 & -12.13  \\
  						& WLC:Session 2 & -0.07 & 0.01 & -7.08 \\
  						& WLC:Session 3	& -0.01 & 0.01 & \textbf{-0.99} \\
\hline
Inverted Words (iW)  	& Intercept 	& 5.48  & 0.04 & 128.67  \\
 						& WLC         	& 0.01  & 0.02 & \textbf{0.45} \\
 						& Session 1     & 0.36  & 0.05 & 7.48  \\
						& Session 2     & 0.27  & 0.03 & 7.84  \\
						& Session 3     & 0.22  & 0.03 & 6.17  \\
						& WLC:Session 1 & -0.21& 0.01& -18.10 \\
  						& WLC:Session 2 & -0.10 & 0.01 & -9.56 \\
 					 	& WLC:Session 3 & -0.06 & 0.01 & -5.71 \\

\hline
Scrambled Letters (sL)  	& Intercept 	& 5.46   & 0.03  & 186.58  \\
  						& WLC         	& 0.01  & 0.02 & \textbf{0.24}  \\
						& Session 1     & 0.11  & 0.04 & 2.91  \\
  						& Session 2     & 0.11  & 0.05 & 2.21  \\
  						& Session 3     & 0.10  & 0.04 & 2.34  \\
 						& WLC:Session 1 & 0.10  & 0.01 & 11.95  \\
 						& WLC:Session 2 & 0.12  & 0.01 & 14.61  \\
 						& WLC:Session 3 & 0.09  & 0.01  & 10.80  \\
\hline
\multicolumn{5}{l}{\tiny{Note: Fixation duration value is log-transformed. WLC = word length. Non-significant values are marked in bold.} }        
\end{tabular}
\end{table}

\newpage

\begin{table}[h]
\footnotesize
\caption{Results from Linear Mixed-Effects model: Duration of First of Multiple Fixations}
\label{LMM_Fmf}
\begin{tabular}{|l|l|c|c|c|}
\hline\hline
Conditions & 	&\textit{b} & \textit{SE} & \textit{t value} \\
\hline
Mirrored Letters (mL)	& Intercept 	& 5.43     & 0.05       & 114.86  \\
 						& WLC			& -0.01    & 0.02       & \textbf{-0.31}   \\
 						& Session 1		& 0.16     & 0.03       & 4.56    \\
 						& Session 2		& 0.04     & 0.03       & \textbf{1.23}    \\
						& Session 3		& 0.03     & 0.03       & \textbf{0.93}    \\
 						& WLC:Session 1	& 0.09     & 0.01       & 6.04    \\
 						& WLC:Session 2	& 0.09     & 0.02       & 5.86    \\
						& WLC:Session 3	& 0.08     & 0.02       & 4.68    \\
\hline
Mirrored Words (mW)  	& Intercept 	& 5.37     & 0.05       & 101.59  \\
						& WLC			& 0.04     & 0.03       & \textbf{1.56}    \\
						& Session 1		& 0.34     & 0.04       & 9.23    \\
						& Session 2		& 0.32     & 0.03       & 10.98   \\
						& Session 3		& 0.24     & 0.05       & 4.36    \\
						& WLC:Session 1 & -0.16    & 0.02       & -10.31  \\
						& WLC:Session 2 & -0.20    & 0.01       & -14.05  \\
						& WLC:Session 3 & -0.10    & 0.01       & -6.60   \\
\hline
Inverted Words (iW)  	& Intercept 	& 5.46     & 0.04       & 122.19  \\
						& WLC			& 0.03     & 0.02       & \textbf{1.22}    \\
						& Session 1 	& 0.33     & 0.05       & 6.32    \\
						& Session 2 	& 0.25     & 0.05       & 4.83    \\
						& Session 3		& 0.20     & 0.05       & 4.14    \\
						& WLC:Session 1	& -0.16    & 0.02       & -10.58  \\
						& WLC:Session 2 & -0.14    & 0.02       & -9.11   \\
						& WLC:Session 3 & -0.11    & 0.02       & -7.32   \\
\hline
Scrambled Letters (sL)  	& Intercept  	& 5.44     & 0.03       & 162.02  \\
						& WLC			& 0.01     & 0.02       & \textbf{0.27}    \\
						& Session 1 	& 0.06     & 0.04       & \textbf{1.54}    \\
						& Session 2		& 0.06     & 0.03       & \textbf{1.84}    \\
						& Session 3		& 0.06     & 0.03       & \textbf{1.77}    \\
						& WLC:Session 1	& 0.15     & 0.02       & 8.80    \\
						& WLC:Session 2 & 0.22     & 0.02       & 12.83   \\
						& WLC:Session 3 & 0.15     & 0.02       & 8.52   \\
\hline
\multicolumn{5}{l}{\tiny{Note: Fixation duration value is log-transformed. WLC = word length. Non-significant values are marked in bold.} }
\end{tabular}
\end{table}

\newpage

\begin{table}[h]
\footnotesize
\caption{Results from Linear Mixed-Effects model: Refixation Probabilities}
\label{LMM_Rfx}
\begin{tabular}{|l|l|c|c|c|c|}
\hline\hline
Condition 				& 	&	\textit{b} & \textit{SE} & \textit{z value} & \textit{Pr(\textgreater{}|z|)} \\
\hline
Mirrored Letters (mL)	& Intercept 	& -1.63    & 0.12       & -13.75  & 0.000                 \\
						& WLC			& 1.15     & 0.09       & 12.52   & 0.000                 \\
						& Session 1		& 1.29     & 0.18       & 7.32    & 0.000                 \\
						& Session 2		& 0.76     & 0.14       & 5.53    & 0.000                 \\
						& Session 3		& 0.48     & 0.17       & 2.75    & 0.006                 \\
						& WLC:Session 1	& 0.37     & 0.05       & 6.86    & 0.000                 \\
						& WLC:Session 2	& 0.41     & 0.06       & 7.33    & 0.000                 \\
						& WLC:Session 3	& 0.38     & 0.06       & 6.48    & 0.000                 \\
\hline
Mirrored Words (mW)  	& Intercept 	& -1.35    & 0.18       & -7.62   & 0.000                 \\
						& WLC 			& 1.15     & 0.11       & 10.50   & 0.000                 \\
						& Session 1		& 1.39     & 0.23       & 6.15    & 0.000                 \\
						& Session 2 	& 1.35     & 0.14       & 9.78    & 0.000                 \\
						& Session 3 	& 1.19     & 0.21       & 5.71    & 0.000                 \\
						& WLC:Session 1 & 0.55     & 0.05       & 10.00   & 0.000                 \\
						& WLC:Session 2 & 0.64     & 0.05       & 12.98   & 0.000                 \\
						& WLC:Session 3 & 0.69     & 0.05       & 13.85   & 0.000                 \\
\hline
Inverted Words (iW)		& Intercept 	& -1.51    & 0.17       & -9.03   & 0.000                 \\
						& WLC			& 1.21     & 0.07       & 18.14   & 0.000                 \\
						& Session 1		& 1.87     & 0.15       & 12.49   & 0.000                 \\
						& Session 2		& 1.69     & 0.17       & 10.06   & 0.000                 \\
						& Session 3		& 1.65     & 0.17       & 9.98    & 0.000                 \\
						& WLC:Session 1	& 0.64     & 0.05       & 11.76   & 0.000                 \\
						& WLC:Session 2 & 0.52     & 0.05       & 9.96    & 0.000                 \\
						& WLC:Session 3 & 0.62     & 0.05       & 12.17   & 0.000                 \\
\hline
Scrambled Letters (sL)	& Intercept 	& -1.51    & 0.15       & -9.73   & 0.000                 \\
						& WLC			& 1.46     & 0.10       & 13.95   & 0.000                 \\
						& Session 1		& 0.78     & 0.23       & 3.48    & 0.000                 \\
						& Session 2		& 0.57     & 0.20       & 2.90    & 0.004                 \\
						& Session 3		& 0.67     & 0.18       & 3.64    & 0.000                 \\
						& WLC:Session 1 & 0.32     & 0.05       & 5.92    & 0.000                 \\
						& WLC:Session 2	& 0.42     & 0.06       & 7.61    & 0.000                 \\
						& WLC:Session 3 & 0.31     & 0.06       & 5.52    & 0.000                 \\
\hline
\multicolumn{6}{l}{\tiny{Note: WLC = word length. Non-significant values are marked in bold.} }
\end{tabular}
\end{table}

\newpage

\begin{table}[h]
\footnotesize
\caption{Results from Linear Mixed-Effects model: Effect sizes accross manipulation types}
\label{LMM_effect}
\begin{tabular}[25,10]{|l|l|c|c|c|c|}
\hline\hline
Measures 	&           & \textit{b} & \textit{SE} & \textit{z value} & \textit{Pr(\textgreater{}|z|)}  \\
\hline
Single   	& Intercept & 5.65     & 0.03       & 213.36  &                       \\
Fixation    & WLC       & -0.03    & 0.01       & -3.57   &                       \\
Duration    & mL     	& 0.09     & 0.02       & 3.95    &                       \\
(SFD)       & sL     	& 0.01     & 0.02       & \textbf{0.46}    &                       \\
		    & mW     	& 0.05     & 0.01       & 3.90    &                       \\
		    & iW     	& 0.04     & 0.01       & 4.07    &                       \\
		    & WLC:mL 	& 0.03     & 0.00       & 6.97    &                       \\
		    & WLC:sL 	& 0.02     & 0.00       & 8.38    &                       \\
		    & WLC:mW 	& -0.05    & 0.00       & -19.63  &                       \\
		    & WLC:iW 	& -0.04    & 0.00       & -19.74  &                       \\
\hline
First of    & Intercept & 5.60     & 0.03       & 212.41  &                       \\
Multiple    & WLC       & 0.00     & 0.01       & \textbf{-0.29}   &                       \\
Fixations   & mL     	& 0.08     & 0.02       & 4.08    &                       \\
duration 	& sL     	& -0.01    & 0.01       & \textbf{-0.42}   &                       \\
(FMF)       & mW     	& 0.06     & 0.01       & 5.47    &                       \\
	        & iW     	& 0.04     & 0.01       & 4.45    &                       \\
	        & WLC:mL 	& 0.04     & 0.01       & 5.82    &                       \\
	        & WLC:sL 	& 0.05     & 0.01       & 9.21    &                       \\
	        & WLC:mW 	& -0.06    & 0.00       & -17.12  &                       \\
	        & WLC:iW 	& -0.04    & 0.00       & -14.35  &                       \\
\hline
Refixation  & Intercept & -0.40    & 0.08       & -5.03   & 0.000                 \\
Probability & WLC       & 1.59     & 0.04       & 36.17   & 0.000                 \\
(RFX)  		& mL     	& 0.62     & 0.08       & 7.33    & 0.000                 \\
		    & sL     	& 0.04     & 0.06       & 0.66    & \textbf{0.506}                 \\
		    & mW     	& 0.21     & 0.04       & 4.80    & 0.000                 \\
		    & iW     	& 0.21     & 0.04       & 5.91    & 0.000                 \\
		    & WLC:mL 	& 0.17     & 0.02       & 7.17    & 0.000                 \\
		    & WLC:sL 	& 0.08     & 0.02       & 4.47    & 0.000                 \\
		    & WLC:mW 	& 0.07     & 0.01       & 5.05    & 0.000                 \\
		    & WLC:iW 	& 0.05     & 0.01       & 4.54    & 0.000                 \\
\hline
\multicolumn{6}{l}{\tiny{Note: LMM model with helmert contrast. SFD and FMD values are log-transformed (base 10).}}\\ 
\multicolumn{6}{l}{\tiny{WLC = word length. Non-significant values are marked in bold.} }
\end{tabular}
\end{table}

\newpage

\begin{table}[p]  
\caption{Quadratic fit to initial landing position curve (two-fixation cases) for reading normal text: Estimates of parameters A, B and C}
\label{Quad_N}
\begin{tabular}{|l|c|c|c|c|c|c|}
\hline
Session & Word Length & Parameters & Estimate & Std. Error & z value & Pr(z) \\
\hline\hline
0       & 4           & A          & 0.051    & 0.020      & 2.466   & 0.013 \\
        & 4           & B          & 0.065    & 0.008      & 7.814   & 0.000 \\
        & 4           & C          & 2.528    & 0.101      & 24.916  & 0.000 \\
        \hline
        & 5           & A          & 0.023    & 0.015      & 1.497   & 0.134 \\
        & 5           & B          & 0.040    & 0.004      & 9.139   & 0.000 \\
        & 5           & C          & 3.336    & 0.121      & 27.468  & 0.000 \\
        \hline
        & 6           & A          & 0.019    & 0.012      & 1.516   & 0.130 \\
        & 6           & B          & 0.023    & 0.003      & 8.859   & 0.000 \\
        & 6           & C          & 4.21     & 0.168      & 25.043  & 0.000 \\
        \hline
        & 7           & A          & -0.008   & 0.016      & -0.518  & 0.604 \\
        & 7           & B          & 0.018    & 0.003      & 7.115   & 0.000 \\
        & 7           & C          & 4.98     & 0.251      & 19.834  & 0.000 \\
        \hline
\end{tabular}
\end{table}  

\newpage 

\begin{table}[p] 
\caption{Quadratic fit to initial landing position curve (two-fixation cases) for reading \textbf{\textit{mirrored-word (mW)}} text: Estimates of parameters A, B and C}
\label{Quad_mW}
\begin{tabular}{|l|c|c|c|c|c|c|}
\hline
Session & Word Length & Parameters & Estimate & Std. Error & z value  & Pr(z) \\
\hline\hline
1       & 4           & A          & 0.080    & 0.022      & 3.684    & 0.000 \\
        & 4           & B          & 0.048    & 0.009      & 5.309    & 0.000 \\
        & 4           & C          & 2.670    & 0.173      & 15.650   & 0.000 \\
\hline
        & 5           & A          & 0.144    & 0.027      & 5.288    & 0.000 \\
        & 5           & B          & 0.007    & 0.007      & 1.037    & 0.300 \\
        & 5           & C          & 2.787    & 0.757      & 3.684    & 0.000 \\
\hline
        & 6           & A          & 0.181    & 0.028      & 6.533    & 0.000 \\
        & 6           & B          & -0.009   & 0.005      & -1.680   & 0.093 \\
        & 6           & C          & 3.523    & 0.602      & 5.852    & 0.000 \\
\hline
        & 7           & A          & 0.173    & 0.019      & 9.009    & 0.000 \\
        & 7           & B          & -0.009   & 0.003      & -3.265   & 0.001 \\
        & 7           & C          & 3.267    & 0.314      & 10.392   & 0.000 \\
\hline \hline
2       & 4           & A          & 0.074    & 0.039      & 1.897    & 0.058 \\
        & 4           & B          & 0.019    & 0.010      & 1.9497   & 0.051 \\
        & 4           & C          & 4.152    & 1.145      & 3.627    & 0.000 \\
\hline
        & 5           & A          & 0.099    & 0.027      & 3.633    & 0.000 \\
        & 5           & B          & 0.012    & 0.008      & 1.518    & 0.129 \\
        & 5           & C          & 4.202    & 1.221      & 3.443    & 0.001 \\
\hline 
        & 6           & A          & 0.056    & 0.086      & 0.650    & 0.516 \\
        & 6           & B          & 0.000    & 0.000      & 1.049    & 0.294 \\
        & 6           & C          & 17.788   & 0.000      & 38076.98 & 0.000 \\
\hline      
        & 7           & A          & 0.166    & 0.025      & 6.692    & 0.000 \\
        & 7           & B          & -0.007   & 0.004      & -2.078   & 0.038 \\
        & 7           & C          & 3.057    & 0.526      & 5.806    & 0.000 \\
\hline\hline
3       & 4           & A          & 0.043    & 0.011      & 3.861    & 0.000 \\
        & 4           & B          & 0.063    & 0.005      & 13.502   & 0.000 \\
        & 4           & C          & 2.718    & 0.069      & 39.444   & 0.000 \\
\hline
        & 5           & A          & 0.0845   & 0.042      & 1.998    & 0.046 \\
        & 5           & B          & 0.011    & 0.009      & 1.119    & 0.263 \\
        & 5           & C          & 4.708    & 2.078      & 2.265    & 0.023 \\
\hline
        & 6           & A          & -0.008   & 0.076      & -0.101   & 0.919 \\
        & 6           & B          & 0.001    & 0.000      & 2.050    & 0.040 \\
        & 6           & C          & 17.752   & 0.001      & 23601.67 & 0.000 \\
\hline 
        & 7           & A          & 0.176    & 0.028      & 6.275    & 0.000 \\
        & 7           & B          & -0.009   & 0.004      & -2.315   & 0.021 \\
        & 7           & C          & 3.101    & 0.465      & 6.666    & 0.000 \\
\hline
\end{tabular}
\end{table} 

\newpage

\begin{table}[p]
\caption{Quadratic fit to initial landing position curve (two-fixation cases) for reading \textbf{\textit{mirrored-letter (mL)}} text: Estimates of parameters A, B and C}
\label{Quad_mL}
\begin{tabular}{|l|c|c|c|c|c|c|}
\hline
Session & Word Length & Parameters & Estimate & Std. Error & z value  & Pr(z) \\
\hline\hline
1       & 4           & A          & -0,007   & 0,022      & -0,314   & 0.753 \\
        & 4           & B          & 0,078    & 0,009      & 8,332    & 0.000 \\
        & 4           & C          & 2,802    & 0,120      & 23,424   & 0.000 \\
\hline
        & 5           & A          & 0,009    & 0,017      & 0,494    & 0.621 \\
        & 5           & B          & 0,038    & 0,005      & 7,626    & 0.000 \\
        & 5           & C          & 3,593    & 0,172      & 20,842   & 0.000 \\
\hline
        & 6           & A          & 0,028    & 0,006      & 4,569    & 0.000 \\
        & 6           & B          & 0,016    & 0,001      & 12,389   & 0.000 \\
        & 6           & C          & 4,800    & 0,161      & 29,770   & 0.000 \\
 \hline
        & 7           & A          & 0,015    & 0,016      & 0,943    & 0.346 \\
        & 7           & B          & 0,007    & 0,002      & 4,283    & 0.000 \\
        & 7           & C          & 6,717    & 0,787      & 8,540    & 0.000 \\
\hline\hline
2       & 4           & A          & 0,002    & 0,004      & 0,446    & 0.656 \\
        & 4           & B          & 0,060    & 0,002      & 36,706   & 0.000 \\
        & 4           & C          & 3,150    & 0,035      & 89,4066  & 0.000 \\
\hline
        & 5           & A          & 1,316    & 0,331      & 3,974    & 0.000 \\
        & 5           & B          & -0,002   & 0,001      & -3,487   & 0.000 \\
        & 5           & C          & -20,033  & 0,018      & -1087,91 & 0.000 \\
\hline
        & 6           & A          & 0,002    & 0,019      & 0,095    & 0.924 \\
        & 6           & B          & 0,022    & 0,004      & 5,463    & 0.000 \\
        & 6           & C          & 4,544    & 0,324      & 14,023   & 0.000 \\
\hline
        & 7           & A          & 0,011    & 0,010      & 1,076    & 0.282 \\
        & 7           & B          & 0,010    & 0,002      & 6,450    & 0.000 \\
        & 7           & C          & 5,933    & 0,405      & 14,661   & 0.000 \\
\hline\hline
3       & 4           & A          & 0,041    & 0,023      & 1,838    & 0.066 \\
        & 4           & B          & 0,054    & 0,010      & 5,436    & 0.000 \\
        & 4           & C          & 2,980    & 0,211      & 14,152   & 0.000 \\
\hline       
        & 5           & A          & 0,010    & 0,021      & 0,487    & 0.631 \\
        & 5           & B          & 0,037    & 0,006      & 6,021    & 0.000 \\
        & 5           & C          & 3,654    & 0,227      & 16,110   & 0.000 \\
\hline
        & 6           & A          & 0,002    & 0,020      & 0,080    & 0.936 \\
        & 6           & B          & 0,024    & 0,004      & 5,568    & 0.000 \\
        & 6           & C          & 4,363    & 0,290      & 15,042   & 0.000 \\
\hline
        & 7           & A          & -0,002   & 0,020      & -0,123   & 0.902 \\
        & 7           & B          & 0,017    & 0,003      & 5,455    & 0.000 \\
        & 7           & C          & 4,990    & 0,329      & 15,169   & 0.000 \\
\hline
\end{tabular}
\end{table} 

\newpage

\begin{table}[p] 
\caption{Quadratic fit to initial landing position curve (two-fixation cases) for \textbf{\textit{reading inverted-word (iW)}} text: Estimates of parameters A, B and C}
\label{Quad_iW}
\begin{tabular}{|l|c|c|c|c|c|c|}
\hline
Session & Word Length & Parameters & Estimate & Std. Error & z value  & Pr(z) \\
\hline\hline
1       & 4           & A          & -0.460   & 0.116      & -3.976   & 0.000 \\
        & 4           & B          & 0.002    & 0.000      & 5.742    & 0.000 \\
        & 4           & C          & 22.504   & 0.004      & 5642.92  & 0.000 \\
\hline        
        & 5           & A          & 0.213    & 0.022      & 9.688    & 0.000 \\
        & 5           & B          & -0.012   & 0.006      & -1.884   & 0.060 \\
        & 5           & C          & 1.510    & 0.653      & 2.312    & 0.021 \\
\hline        
        & 6           & A          & 0.214    & 0.026      & 8.132    & 0.000 \\
        & 6           & B          & -0.018   & 0.005      & -3.570   & 0.000 \\
        & 6           & C          & 2.940    & 0.243      & 12.093   & 0.000 \\
\hline        
        & 7           & A          & 0.197    & 0.023      & 8.550    & 0.000 \\
        & 7           & B          & -0.013   & 0.003      & -3.954   & 0.000 \\
        & 7           & C          & 3.064    & 0.276      & 11.102   & 0.000 \\
\hline\hline
2       & 4           & A          & 0.058    & 0.014      & 4.153    & 0.000 \\
        & 4           & B          & 0.030    & 0.005      & 5.758    & 0.000 \\
        & 4           & C          & 3.677    & 0.309      & 11.908   & 0.000 \\
 \hline
        & 5           & A          & 0.006    & 0.054      & 0.115    & 0.909 \\
        & 5           & B          & 0.002    & 0.001      & 3.198    & 0.001 \\
        & 5           & C          & 11.017   & 0.001      & 11956.68 & 0.000 \\
\hline
        & 6           & A          & 0.042    & 0.116      & 0.361    & 0.718 \\
        & 6           & B          & 0.000    & 0.000      & 0.881    & 0.378 \\
        & 6           & C          & 24.665   & 0.001      & 41339.44 & 0.000 \\
\hline
        & 7           & A          & 0.167    & 0.025      & 6.785    & 0.000 \\
        & 7           & B          & -0.008   & 0.004      & -2.279   & 0.000 \\
        & 7           & C          & 3.522    & 0.439      & 8.024    & 0.000 \\
\hline\hline
3       & 4           & A          & 0.021    & 0.010      & 2.219    & 0.000 \\
        & 4           & B          & 0.053    & 0.004      & 12.494   & 0.000 \\
        & 4           & C          & 3.185    & 0.106      & 30.049   & 0.000 \\
\hline 
        & 5           & A          & 0.080    & 0.029      & 2.767    & 0.006 \\
        & 5           & B          & 0.014    & 0.008      & 1.763    & 0.078 \\
        & 5           & C          & 4.295    & 1.099      & 3.908    & 0.000 \\
\hline        
        & 6           & A          & 0.057    & 0.114      & 0.502    & 0.616 \\
        & 6           & B          & 0.000    & 0.000      & 0.767    & 0.443 \\
        & 6           & C          & 25.887   & 0.000      & 51952.32 & 0.000 \\
\hline   
        & 7           & A          & 0.165    & 0.022      & 7.515    & 0.000 \\
        & 7           & B          & -0.008   & 0.003      & -2.388   & 0.017 \\
        & 7           & C          & 3.685    & 0.426      & 8.654    & 0.000 \\
\hline
\end{tabular}
\end{table} 

\newpage
\begin{table}[p] 
\caption{Quadratic fit to initial landing position curve (two-fixation cases) for reading \textbf{\textit{scrambled-letter (sL)}} text: Estimates of parameters A, B and C}
\label{Quad_sL}
\begin{tabular}{|l|c|c|c|c|c|c|}
\hline
Session                & Word Length & Parameters & Estimate & Std. Error & z value   & Pr(z) \\
\hline\hline
1                      & 4           & A          & -0.034   & 0.034      & -1.020    & 0.301 \\
                       & 4           & B          & 0.087    & 0.014      & 6.051     & 0.000 \\
                       & 4           & C          & 2.833    & 0.169      & 16.776    & 0.000 \\
\hline                     
                       & 5           & A          & 0.017    & 0.008      & 2.123     & 0.034 \\
                       & 5           & B          & 0.031    & 0.0024     & 13.118    & 0.000 \\
                       & 5           & C          & 3.861    & 0.118      & 32.791    & 0.000 \\
\hline
                       & 6           & A          & 0.022    & 0.008      & 2.859     & 0.004 \\
                       & 6           & B          & 0.019    & 0.002      & 11.561    & 0.000 \\
                       & 6           & C          & 4.563    & 0.155      & 29.519    & 0.000 \\
\hline
                       & 7           & A          & 0.620    & 0.043      & 14.332    & 0.000 \\
                       & 7           & B          & 0.000    & 7.177      & -11.598   & 0.000 \\
                       & 7           & C          & -20.736  & 0.001      & -19706.55 & 0.000 \\
\hline\hline
2                      & 4           & A          & 0.002    & 0.026      & 0.081     & 0.935 \\
                       & 4           & B          & 0.086    & 0.010      & 8.373     & 0.000 \\
                       & 4           & C          & 2.538    & 0.095      & 26.578    & 0.000 \\
\hline
                       & 5           & A          & -0.009   & 0.024      & -0.363    & 0.716 \\
                       & 5           & B          & 0.042    & 0.007      & 6.196     & 0.000 \\
                       & 5           & C          & 3.580    & 0.210      & 17.015    & 0.000 \\
\hline
                       & 6           & A          & 0.021    & 0.015      & 1.348     & 0.178 \\
                       & 6           & B          & 0.014    & 0.003      & 4.705     & 0.000 \\
                       & 6           & C          & 5.149    & 0.492      & 10.456    & 0.000 \\
\hline
                       & 7           & A          & 0.618    & 0.055      & 11.303    & 0.000 \\
                       & 7           & B          & -0.001   & 0.000      & -9.141    & 0.000 \\
                       & 7           & C          & -20.966  & 0.001      & -17086.51 & 0.000 \\
\hline\hline
3                      & 4           & A          & 1.376    & 0.758      & 1.815     & 0.070 \\
                       & 4           & B          & -0.002   & 0.001      & -1.553    & 0.121 \\
                       & 4           & C          & -22.215  & 0.041      & -547.09   & 0.000 \\
\hline
                       & 5           & A          & 0.025    & 0.009      & 2.778     & 0.005 \\
                       & 5           & B          & 0.029    & 0.003      & 10.999    & 0.000 \\
                       & 5           & C          & 3.918    & 0.145      & 27.020    & 0.000 \\
\hline
                       & 6           & A          & 0.020    & 0.020      & 0.984     & 0.325 \\
                       & 6           & B          & 0.015    & 0.004      & 3.698     & 0.000 \\
                       & 6           & C          & 5.046    & 0.601      & 8.399     & 0.000 \\
\hline
                       & 7           & A          & -0.027   & 0.011      & -2.442    & 0.015 \\
                       & 7           & B          & 0.003    & 0.000      & 16.467    & 0.000 \\
                       & 7           & C          & 10.061   & 0.000      & 49984.09  & 0.000 \\
\hline
\end{tabular}
\end{table} 

\begin{figure}[t]
\unitlength1mm
\begin{picture}(150,150)
\put(-3,-10){\includegraphics[scale=.80]{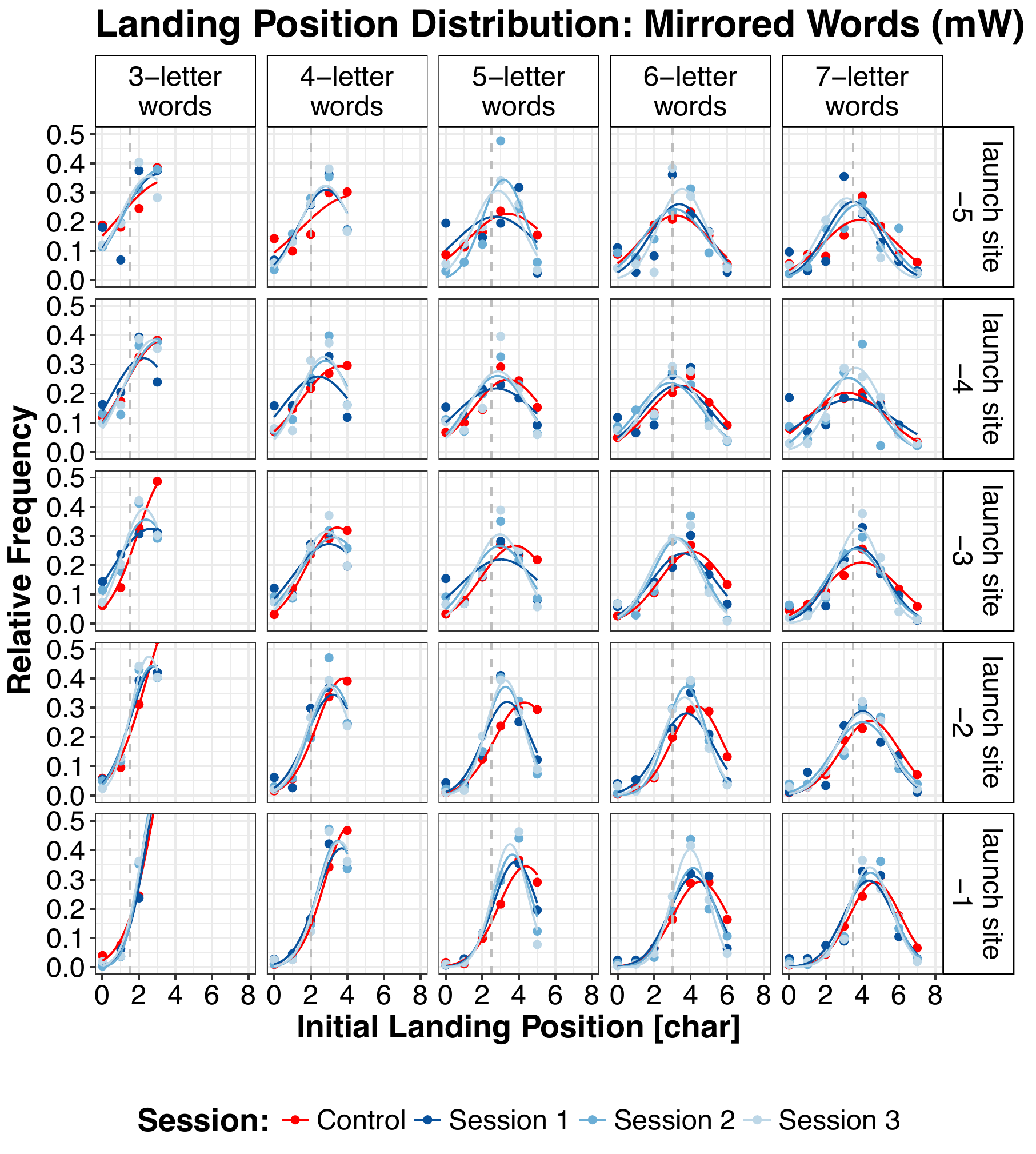}}
\end{picture}
\vspace{10mm}
\caption{\label{landpos_mW}
Within-word landing position distribution group by launch-site distance and word length for reading \textit{mirrored-word (mW)} texts. Red line and dots represent data from normal reading session. Data from the first experimental session are presented in dark blue color. The lighter blue hues represent the last two experimental sessions. }
\end{figure}

\begin{figure}[t]
\unitlength1mm
\begin{picture}(150,150)
\put(-3,-10){\includegraphics[scale=.80]{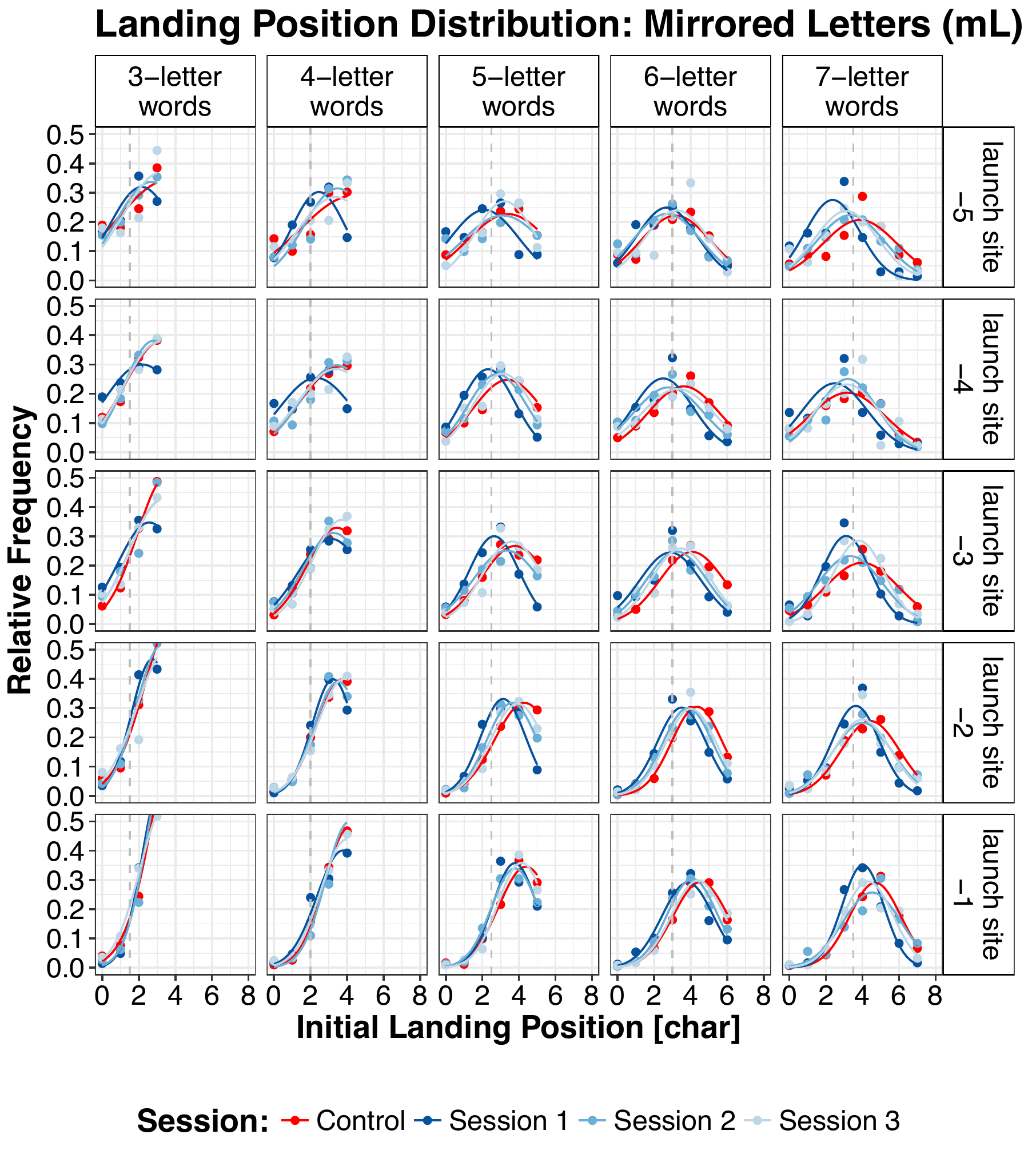}}
\end{picture}
\vspace{10mm}
\caption{\label{landpos_mL}
Within-word landing position distribution group by launch-site distance and word length for reading \textit{mirrored-letter (mL)} texts. Red line and dots represent data from normal reading session. Data from the first experimental session are presented in dark blue color. The lighter blue hues represent the last two experimental sessions. }
\end{figure}

\begin{figure}[t]
\unitlength1mm
\begin{picture}(150,150)
\put(-3,-10){\includegraphics[scale=.80]{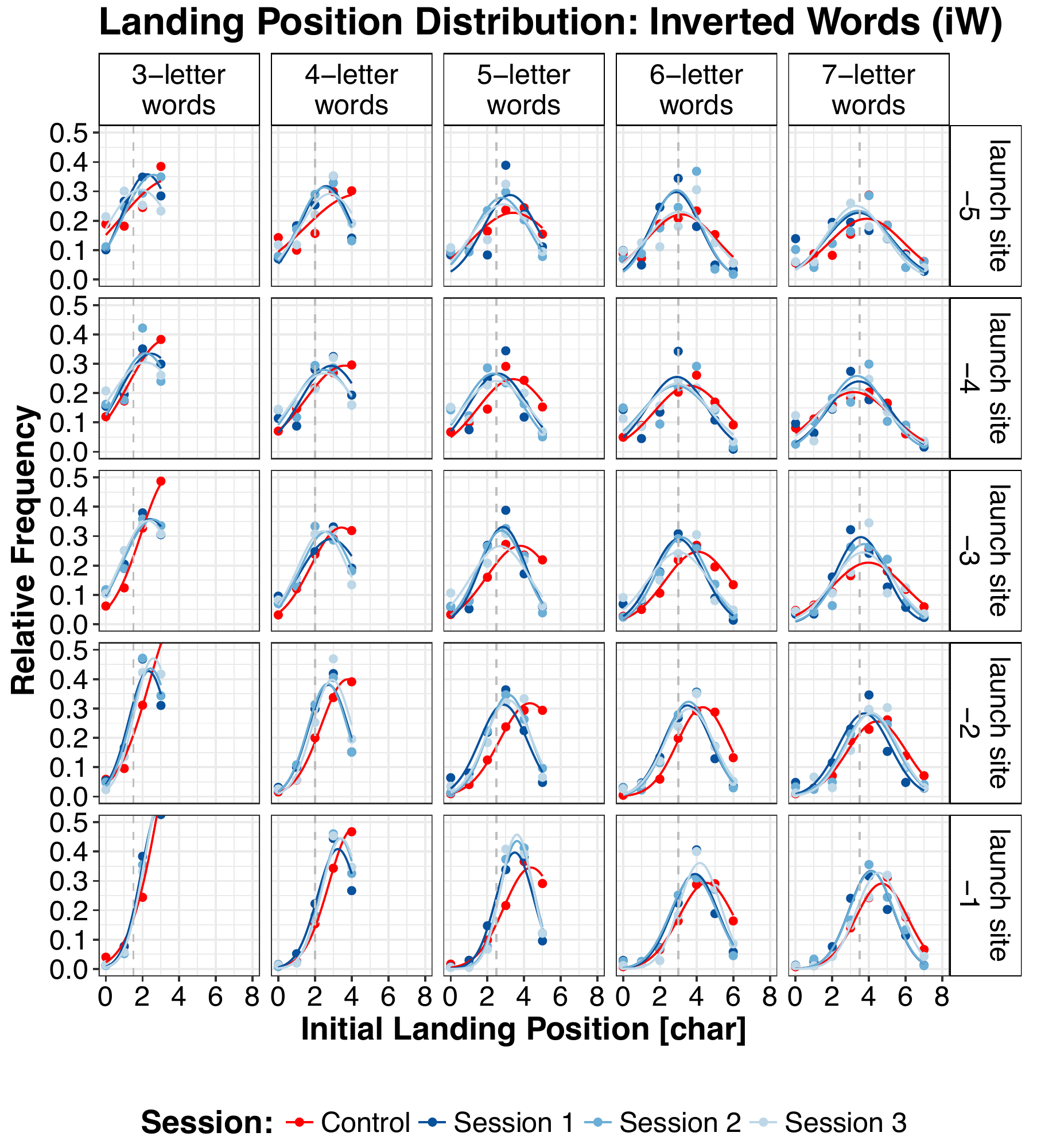}}
\end{picture}
\vspace{10mm}
\caption{\label{landpos_iW}
Within-word landing position distribution group by launch-site distance and word length for reading \textit{inverted-word (iW)} texts. Red line and dots represent data from normal reading session. Data from the first experimental session are presented in dark blue color. The lighter blue hues represent the last two experimental sessions. }
\end{figure}

\begin{figure}[t]
\unitlength1mm
\begin{picture}(150,150)
\put(-3,-10){\includegraphics[scale=.80]{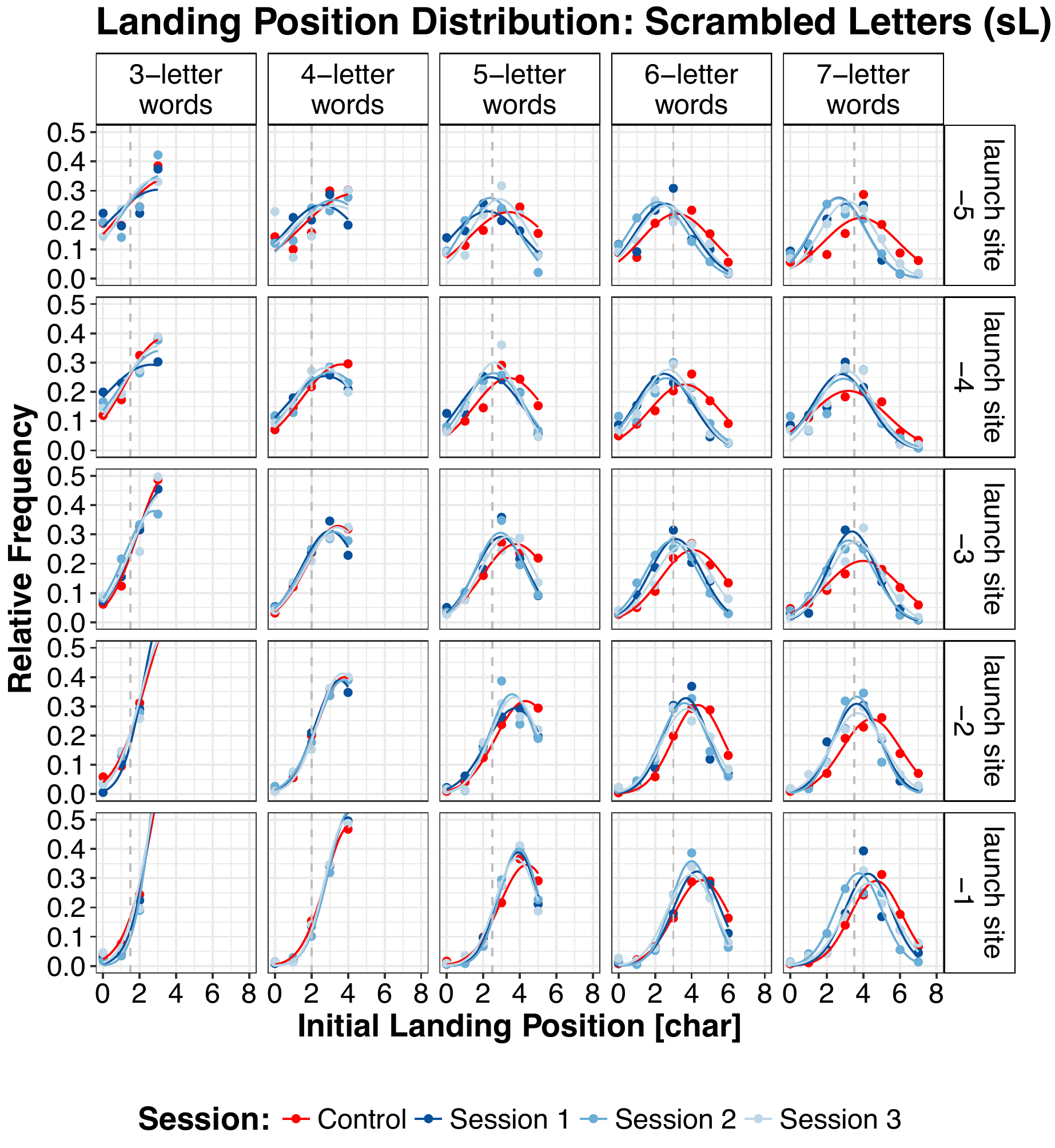}}
\end{picture}
\vspace{10mm}
\caption{\label{landpos_sL}
Within-word landing position distribution group by launch-site distance and word length for reading \textit{scrambled-letter (sL)} texts. Red line and dots represent data from normal reading session. Data from the first experimental session are presented in dark blue color. The lighter blue hues represent the last two experimental sessions. }
\end{figure}

\end{appendix}

\end{document}